\documentclass[nonacm]{acmart}
\newcommand{\rev}[1]{#1}

\usepackage{fancybox}
\usepackage{listings}
\usepackage{enumitem}
\usepackage{adjustbox} 
\usepackage{color,soul}%for usage of style attributes - background color
\usepackage{tikz}
\usepackage{xargs}
\usepackage[colorinlistoftodos,prependcaption,textsize=tiny]{todonotes}
\newcommandx{\unsure}[2][1=]{\todo[linecolor=red,backgroundcolor=red!25,bordercolor=red,#1]{#2}}
\newcommandx{\change}[2][1=]{\todo[linecolor=blue,backgroundcolor=blue!25,bordercolor=blue,#1]{#2}}
\newcommandx{\info}[2][1=]{\todo[linecolor=OliveGreen,backgroundcolor=OliveGreen!25,bordercolor=OliveGreen,#1]{#2}}
\newcommandx{\improvement}[2][1=]{\todo[linecolor=Plum,backgroundcolor=Plum!25,bordercolor=Plum,#1]{#2}}
\newcommandx{\thiswillnotshow}[2][1=]{\todo[disable,#1]{#2}}

\usetikzlibrary{calc}
\usetikzlibrary{decorations.pathmorphing}

\makeatletter

\newcommand{\defhighlighter}[3][]{%
  \tikzset{every highlighter/.style={color=#2, fill opacity=#3, #1}}%
}

\defhighlighter{yellow}{.5}

\newcommand{\highlight@DoHighlight}{
  \fill [ decoration = {random steps, amplitude=1pt, segment length=15pt}
        , outer sep = -15pt, inner sep = 0pt, decorate
        , every highlighter, this highlighter ]
        ($(begin highlight)+(0,8pt)$) rectangle ($(end highlight)+(0,-3pt)$) ;
}

\newcommand{\highlight@BeginHighlight}{
  \coordinate (begin highlight) at (0,0) ;
}

\newcommand{\highlight@EndHighlight}{
  \coordinate (end highlight) at (0,0) ;
}

\newdimen\highlight@previous
\newdimen\highlight@current

\DeclareRobustCommand*\highlight[1][]{%
  \tikzset{this highlighter/.style={#1}}%
  \SOUL@setup
  \def\SOUL@preamble{%
    \begin{tikzpicture}[overlay, remember picture]
      \highlight@BeginHighlight
      \highlight@EndHighlight
    \end{tikzpicture}%
  }%
  \def\SOUL@postamble{%
    \begin{tikzpicture}[overlay, remember picture]
      \highlight@EndHighlight
      \highlight@DoHighlight
    \end{tikzpicture}%
  }%
  \def\SOUL@everyhyphen{%
    \discretionary{%
      \SOUL@setkern\SOUL@hyphkern
      \SOUL@sethyphenchar
      \tikz[overlay, remember picture] \highlight@EndHighlight ;%
    }{%
    }{%
      \SOUL@setkern\SOUL@charkern
    }%
  }%
  \def\SOUL@everyexhyphen##1{%
    \SOUL@setkern\SOUL@hyphkern
    \hbox{##1}%
    \discretionary{%
      \tikz[overlay, remember picture] \highlight@EndHighlight ;%
    }{%
    }{%
      \SOUL@setkern\SOUL@charkern
    }%
  }%
  \def\SOUL@everysyllable{%
    \begin{tikzpicture}[overlay, remember picture]
      \path let \p0 = (begin highlight), \p1 = (0,0) in \pgfextra
        \global\highlight@previous=\y0
        \global\highlight@current =\y1
      \endpgfextra (0,0) ;
      \ifdim\highlight@current < \highlight@previous
        \highlight@DoHighlight
        \highlight@BeginHighlight
      \fi
    \end{tikzpicture}%
    \the\SOUL@syllable
    \tikz[overlay, remember picture] \highlight@EndHighlight ;%
  }%
  \SOUL@
}

\newcommand{\name}[1]{\noindent\hl{\texttt{#1}}}

\newcommand{\just}[1]{\textcolor{blue!70}{#1}}

\AtBeginDocument{%
  \providecommand\BibTeX{{%
    \normalfont B\kern-0.5em{\scshape i\kern-0.25em b}\kern-0.8em\TeX}}}

\setcopyright{acmcopyright}
\copyrightyear{2018}
\acmYear{2018}
\acmDOI{1}

\acmConference[Woodstock '18]{Woodstock '18: ACM Symposium on Neural
  Gaze Detection}{June 03--05, 2018}{Woodstock, NY}
\acmBooktitle{Woodstock '18: ACM Symposium on Neural Gaze Detection}
\acmPrice{15.00}
\acmISBN{978-1-4503-XXXX-X/18/06}

\begin{document}

%%
%% The "title" command has an optional parameter,
%% allowing the author to define a "short title" to be used in page headers.
\title{Facilitating Knowledge Sharing from Domain Experts to Data Scientists for Building NLP Models}

%  Eliciting Domain Knowledge from Domain Experts for Cold Start Scenarios in NLP domain 
% 

%%
%% The "author" command and its associated commands are used to define
%% the authors and their affiliations.
%% Of note is the shared affiliation of the first two authors, and the
%% "authornote" and "authornotemark" commands
%% used to denote shared contribution to the research.
\author{Soya Park}
\email{soya@mit.edu}
\affiliation{%
  \institution{Massachusetts Institute of Technology}
  \country{USA}
}

\author{April Wang}
\affiliation{%
  \institution{University of Michigan}
  \country{USA}}
\email{aprilww@umich.edu}

\author{Ban Kawas}
\author{Q. Vera Liao}
\author{David Piorkowski}
\author{Marina Danilevsky}
\email{bkawas@us.ibm.com}
\email{vera.liao@ibm.com}
\email{david.piorkowski@ibm.com}
\email{mdanile@us.ibm.com}
\affiliation{%
  \institution{IBM}
  \country{USA}
}

%%
%% By default, the full list of authors will be used in the page
%% headers. Often, this list is too long, and will overlap
%% other information printed in the page headers. This command allows
%% the author to define a more concise list
%% of authors' names for this purpose.
\renewcommand{\shortauthors}{S. Park et al.}

%%
%% The abstract is a short summary of the work to be presented in the
%% article.
\begin{abstract}
Data scientists face a steep learning curve in understanding a new domain for which they want to build machine learning (ML) models. While input from domain experts could offer valuable help, such input is often limited, expensive, and generally not in a form readily consumable by a model development pipeline. In this paper, we propose Ziva, a framework to guide domain experts in sharing essential domain knowledge to data scientists for building NLP models. 
% Ziva provides
With Ziva, experts are able to distill and share their domain knowledge using domain concept extractors and five types of label justification over a representative data sample. %Ziva is especially useful in cold-start situations (no training data available), and lowers the communication barrier between domain experts and data scientists. 
The design of Ziva is informed by preliminary interviews with data scientists, in order to understand current practices of domain knowledge acquisition process for ML development projects. To assess our design, we run a mix-method case-study to evaluate how Ziva can facilitate interaction of domain experts and data scientists. Our results highlight that (1) domain experts are able to use Ziva to provide rich domain knowledge, while maintaining low mental load and stress levels;  and (2) data scientists find Ziva's output helpful for learning essential information about the domain, offering scalability of information, and lowering the burden on domain experts to share knowledge. We conclude this work by experimenting with building  NLP models using the Ziva output by our case study.

\end{abstract}

%%
%% The code below is generated by the tool at http://dl.acm.org/ccs.cfm.
%% Please copy and paste the code instead of the example below.
%%


%%
%% Keywords. The author(s) should pick words that accurately describe
%% the work being presented. Separate the keywords with commas.

%% A "teaser" image appears between the author and affiliation
%% information and the body of the document, and typically spans the
%% page.
% \begin{teaserfigure}
%   \includegraphics[width=\textwidth]{sampleteaser}
%   \caption{Seattle Mariners at Spring Training, 2010.}
%   \Description{Enjoying the baseball game from the third-base
%   seats. Ichiro Suzuki preparing to bat.}
%   \label{fig:teaser}
% \end{teaserfigure}

%%
%% This command processes the author and affiliation and title
%% information and builds the first part of the formatted document.
\maketitle

\section{Introduction}

In recent decades, machine learning (ML) technologies are sought by an increasing number of professionals to automate their work tasks or augment their decision-making~\cite{yang2019unremarkable}. Broad areas of applications are benefiting from integration of ML, such as healthcare~\cite{48431, 47823},  finance~\cite{culkin2017machine}, employment~\cite{manyika2017future}, and so on. However, building an ML model in a specialized domain is still expensive and time-consuming for at least two reasons. First, a common bottleneck in developing modern ML technologies is the requirement of a large quantity of labeled data. Second, many steps in \rev{an} ML development pipeline, from problem definition to feature engineering to model debugging, necessitate an understanding of domain-specific knowledge and requirements. %For data scientists assigned to a modeling task, it is challenging for them to establish such an understanding within a limited project period. 
Data scientists therefore often require input from domain experts to both obtain labeled data, as well as understand model requirements, inspire feature engineering, and get feedback on model behavior. In practice, such knowledge transfer between domain experts and data scientists is very much ad-hoc, with no standardized practices or proven effective approaches, and requires significant direct interaction between data scientists and domain experts. 
Building a high-quality legal, medical, or financial classifier will inevitably require a data scientist to consult with professionals in such domains. In practice these are often costly and frustrating iterative conversations and labeling exercises that can go on for weeks and months, which usually still do not yield output in a form readily consumable by a model development pipeline.
% as both sides struggle to understand each other.
%For example, if one data scientist were to build an ML model in legal domain, they first need to understand the basic concepts of the domain... \SP{@marina, can you write down your struggle \& example here}
%These processes could take multiple months to even years. 
% For example, domain experts may ignore details or basis knowledge due to expert blind spot, while data scientists may encounter the unknown of unknowns with not knowing what to ask. 
% \AW{propose adding an example here}
%Ineffective or inefficient knowledge sharing both poses challenges for data scientists to develop models, and result in time loss and frustration for domain experts.

%because it requires data scientists a substantial amount of domain knowledge~\cite{XX}. This makes data scientists heavily rely on domain experts during the ML development, such as labeling additional datasets~\cite{48901} and debugging models~\cite{katsis2019modellens,pinhanez2019machine}.

In this work, we set out to develop methods and interfaces that facilitate knowledge sharing from domain experts to data scientists for model development. We chose to focus on natural language processing (NLP) modeling tasks, and we are especially motivated by real-world cold-start scenarios where labeled data is small or nonexistent. Informed by a formative interview with data scientists regarding current practices and challenges of learning from domain experts, we developed a domain-knowledge acquisition interface \textbf{Ziva} (With \textbf{Z}ero knowledge, How \textbf{I} do de\textbf{V}elop \textbf{A} machine learning model?). Instead of a data-labeling tool, Ziva intends to provide a diverse set of elicitation methods to gather knowledge from domain experts, then present the results as a repository to data scientists to serve their domain understanding needs and to build ML models for specialized domains. Ziva scaffolds the knowledge sharing in desired formats and allows asynchronous exchange between domain experts and data scientists. It also allows flexible re-use of the knowledge repository for different modeling tasks in the domain. 

Specifically, informed by findings from the formative interview and requirements of NLP modeling tasks, Ziva focuses on eliciting key concepts in the text data of a domain (\name{concept creation}), and rationale justifying a label that a domain expert gives to a representative data instance  (\name{justification elicitation}). In the current version of Ziva, we provide five different \name{justification elicitation} methods -- \just{bag of words}, \just{simplification}, 
\just{perturbation}, \just{concept bag of words}, and \just{concept annotation}.

%To devise efficient domain knowledge acquisition, we examined the emergent practices of domain knowledge sharing from domain experts to data scientists and system designs that could facilitate the sharing between two parties. From a formative interview with data scientists, we learned that current practices and challenges of working with domain experts to build an ML model in specialized domains.  

%Informed by this, we developed a domain-knowledge acquisition interface \textbf{Ziva} (With Zero knowledge, How I do deVelop A machine learning model?). Ziva provides interfaces that domain experts can extract basic concepts and components of a domain (\name{concept creation}) and explain rationales of labeling of representative inputs (\name{justification elicitation}). Reflecting our findings from the interview study, we devise 5 different \name{justification elicitation} methods -- \just{bag of words}, \just{simplification}, 

To evaluate and inform future development of Ziva, we conducted a case study in assessment of its coupled design goals: 1) to provide an efficient and user-friendly experience for domain experts to supply domain knowledge; 2) to support data scientists building NLP models, especially in cold-start scenarios.

% To evaluate the Ziva interface for domain experts, 
We conducted a lab study (N=12) and a crowd-deployment study (N=88) for participants to act as domain experts of a restaurant reviewing domain, and use Ziva to provide concepts and justification-based knowledge.  We found the completion time and subjective workload using different elicitation methods varied. Interestingly, the popular keywords based justification (\just{bag of words}) approach led to higher self-reported task success but was considered more stressful. 

We conducted \rev{an interview study} with 7 data scientists to investigate whether and how Ziva could help them build NLP models. Through the study, we identified design requirements for domain knowledge-sharing tools in ML development workflow -- scalability of information and lowering workload for domain experts.
Participants also reflected on how the shared domain knowledge facilitated by Ziva may be utilized, including bootstrapping labels, supporting feature engineering, improving explainability, and training few-shot learning models. Based on these suggestions, we experimented with building a rule-based model using the data from our user study, and report the outcomes using knowledge elicited with different methods.
%Their feedback validated Ziva in facilitating knowledge sharing and better collaboration between domain experts and data scientists, and suggested design requirements for tools to serve these purposes, including scalability and lowering workload for domain experts. %\VL{will come back after finalizing the qual result}. Inspired by the feedback, we experimented with building a rule-based model and a few-shot learning model using the domain knowledge data collected by Ziva. \AW{is this model experimented by the research team or the 7 data scientists?}Reflecting a contextual inquiry, we built a rule-based model to bootstrap labels using the output created during the crowd experiments. As a result, XX \SP{marina results go here.. }
%\VL{need to come back after figuring out the stats}.
%With data generated by domain experts from Ziva, we interviewed data scientists who have worked on building ML models in specialized domains. We found that XX...
%To assess user experiences of domain experts using Ziva, we conducted a lab study (N=12), followed by a crowd-deployment study to focus on comparing \name{justification elicitation} interfaces (N=88). We found that domain experts of basic keyword based labeling, \just{bag of words} self-reported higher success in task accomplishment than other 4 justification techniques, however, they were more stressed during the tasks. 
%Using justification datasets, we built a ruled-based model and few-shot learning model 
%as our interviewee suggested. As a result, we showed XX..  
In summary, the contributions of the paper are as follows:
\begin{itemize}[noitemsep,topsep=0pt]
\item Through a formative interview with data scientists who built models in a specialized domain, we identified their under-supported needs to learn about a domain from domain experts.  

\item We developed Ziva, a tool providing \name{concept creation} and five kinds of \name{justification elicitation} interfaces to gather domain knowledge from domain experts in formats that could help data scientists build NLP models.

\item We conducted a case study using Ziva to elicit domain knowledge then presented the output to data scientists in \rev{an interview} study. Their feedback validated the utility of Ziva and provided design insights for tools that support knowledge sharing and collaboration between domain experts and data scientists.

%An interview study of data scientists using the Ziva output. We identified what are the important measure and desire in domain knowledge sharing. Data scientists picked scalability of the information and having less burden on domain experts is important criteria.    

\item  We also investigated the experience of domain experts using Ziva. We believe that our analysis could inform the design of knowledge elicitation methods for domain experts. 
\end{itemize}

\section{Related work}

We are informed by recent studies of data science practices, as well as ML and HCI work that leverages domain experts' input to train or improve models, and research to facilitate knowledge sharing in teams and organizations.

\subsection{Data Science practices and collaboration}
% Recently the data science domain has spurred great research interest in the HCI community. Besides developing numerous tools to support specific data science tasks (e.g.~\cite{hohman2018visual,hohman2019gamut,zhang2018manifold,ackerman1998augmenting}), an emerging area of research has focused on studying the practices of data scientists in model development work. For example, Muller et al. interviewed data scientists in enterprises and described their active involvement in acquiring, creating, curating and shaping their data, often relying on ``an intuitive sense of their data and processes''~\cite{muller2019datascience}. Many recognized the collaborative nature of data science projects. For example, Kim et al. described five working styles of data scientists in a ML project, including Team Leaders who actively promote and enable early collaboration with customers and other stakeholders~\cite{kim2016emerging}. Zhang et al. described the general collaboration practices of data science teams with a variety of stakeholders ~\cite{zhang2020data}, including engineers, domain experts, managers and communicators. In particular, they showed that domain experts are often actively engaged in consultation during core modeling building stages, such as data-access and feature-extraction. They also take prominent roles in later stages of data science projects for model evaluation and communication of results. 

Recently the data science domain has spurred great research interest in the HCI community. Besides developing numerous tools to support specific data science tasks (e.g.~\cite{hohman2018visual,hohman2019gamut,zhang2018manifold,ackerman1998augmenting}), an emerging area of research has focused on studying the practices of data scientists in model development work. Many have recognized the collaborative nature of data science projects, with both intra- (among data scientists)~\cite{kim2016emerging} and multi-disciplinary (with domain experts) collaboration~\cite{zhang2020data}. In particular, data scientists rely heavily on domain experts during core modeling building stages, such as data access and feature extraction. Domain experts also feature prominently in latter stages of data science projects such as model evaluation and communication of results. However, data scientists' work faces significant challenges as such collaborative activities are currently not well supported~\cite{passi2018trust, mao2019data}, and they are often left with no choice but to rely on ``an intuitive sense of their data and processes''~\cite{muller2019datascience}. %Passi and Jackson highlight that the challenged collaboration not only results in tension in the data science project, but also produces models that lack credibility and can be harmful for the downstream tasks~\cite{passi2018trust}. Mao et al. examined the breakdowns in collaboration between data scientists and domain experts, and identified problems in collaboration readiness, technology readiness, coupling of work dimensions, and tensions that exist in the common ground building process ~\cite{mao2019data}.

% Data scientists' work faces significant challenges as the collaborative activities are currently not well supported. Passi and Jackson highlights that the challenged collaboration not only results in tension in the data science project, but also produces models that lack credibility and can be harmful for the downstream tasks~\cite{passi2018trust}. Mao et al. examined the breakdowns in collaboration between data scientists and domain experts, and identified problems in collaboration readiness, technology readiness, coupling of work dimensions, and tensions that exist in the common ground building process ~\cite{mao2019data}. 
Computational notebooks are positioned as a potential solution to both support collaborative coding and communicating results to stakeholders~\cite{wang2019data}. However, a recent study reported reluctance for data scientists to directly communicate the in-progress model work in notebooks~\cite{rule2018aiding}. While there are tools emerging to address the technology gaps to support collaborative data science practices ~\cite{wang2020callisto,head2019managing,crowston2019socio}, to our knowledge they tend to focus on supporting teams of data scientists and place domain experts with limited elicitation. %a limited utterances. 
%in a passive role. 
In this work, we explore the approach of providing interfaces in which domain experts can create a knowledge repository for a sophisticated domain, so that it could be consumed by data scientists asynchronously and flexibly when the availability of domain experts is limited.

\subsection{ML with domain experts} 
There has been a long-standing desire to increase the involvement of domain experts in model building in both the ML and HCI communities. 
\rev{For example, tasks like text annotation, image annotation involve massive input from domain experts to provide domain-related feedback.
Ziva interface is inspired by NLP text annotation tools like Doccano~\cite{doccano} or Prodigy~\cite{prodigy}, which ease the burden of manual labeling by various visual designs (e.g., using colors to highlight different entities).
We take this design further to acquire domain knowledge for model development and data scientists.
}
Recognizing the challenge of having domain experts label a large quantity of data, many ML works have explored more efficient learning algorithms to reduce the workload~\cite{snell2017prototypical,settles2009active}, or utilize domain experts input as rules~\cite{krishnamurthy2009systemt},  constraints~\cite{chang2007guiding,niculescu2006bayesian}, prior information~\cite{andrzejewski2009incorporating,simard2017machine}, or feedback to re-weigh features~\cite{raghavan2006active,druck2009active}. Recently, given the prominence of label-hungry ML algorithms, weak supervision has become a popular approach to bootstrap labels based on a small amount of labeled input from domain experts~\cite{dehghani2017neural,ratner2017snorkel,ratner2016data}.

The HCI community is further concerned with the isolation of domain experts from the model development process, requiring thus data scientists to go through lengthy and asynchronous iterations to get their input~\cite{amershi2014power, pinhanez2019machine}. To tackle the problem, the sub-field of interactive machine learning (iML) is motivated to enable domain experts or end-users to directly drive model behaviors~\cite{amershi2014power,ware2001interactive,holzinger2016interactive}. Since domain experts might not have training in ML or programming, iML systems elicit their input through intuitive and interactive interfaces (e.g., visualization~\cite{jiang2019recent}, graphic user interfaces~\cite{amershi2014power}, conversational interfaces~\cite{cakmak2012designing}), and a tight feedback loop for them to adjust their input. A variety of user input have been explored in prior work for different tasks in model development, including unseen training data to help correct the model's mistakes~\cite{cakmak2012designing,fails2003interactive}, provide new feature-level input~\cite{krause2014infuse} or adjustment to feature weights~\cite{settles2009active}, assessment of model performance~\cite{fogarty2008cueflik,amershi2015modeltracker}, error preferences~\cite{kapoor2010interactive}, parameter choices~\cite{muhlbacher2017treepod}, model ensemble~\cite{talbot2009ensemblematrix}, etc.

Research on iML has been especially fruitful for NLP modeling tasks, partly because text data and features (e.g., bag of words) are often comprehensible to people without ML training, increasing the likelihood of obtaining effective feedback from domain experts or end-users. For example, for document classification task, DUALIST~\cite{settles2011closing} solicits feedback for both labels and learned features (keywords that the model believes to be informative of the target class). FeatureInsight~\cite{brooks2015featureinsight} further supports feature ideation by the users, e.g. by adding words or creating dictionaries that the user believes to be informative of the target class. EluciDebug~\cite{kulesza2015principles} is an end-user debugging tool that allows critiquing model weights based on the model's explanation on how it classifies a document.
Interactive topic modeling is another well-explored area to incorporate domain experts' input~\cite{choo2013utopian,hoque2015convisit,hu2014interactive,smith2018closing}, for example, by moving documents around or adding words, to refine clusters of topics.

Our work is informed by prior work on iML but takes a complementary approach by facilitating knowledge sharing from domain experts to data scientists. iML is not a panacea to effectively leverage domain experts' input. There are known issues with letting ML novices directly adjust models~\cite{smith2020digging}, such as lacking generalization or over-fitting~\cite{wu2019local}. In practice it is not always feasible to set up an iML system for domain experts to work with. Currently most ML projects still rely on data scientists to write code and set up the pipelines\rev{~\cite{piorkowski2021ai}}. Moreover, having data scientists mediate the knowledge input offers the flexibility to apply it to different kinds of ML algorithms, and allow domain experts to provide reusable knowledge not constrained by a particular modeling task.

In general, it is possible to elicit diverse kinds of knowledge from people, not all of which could be consumed directly by a given ML model. For example, Stumpt et al. ~\cite{stumpf2007toward} and Ghai et al.~\cite{ghai2020explainable} explored what kind of feedback people naturally want to give seeing model explanations. Only a small subset of the various forms of feedback is readily consumable by existing ML algorithms. However, as the ML field rapidly advances, many novel usages of domain knowledge are being explored. For example, since ML models might use low-level features that are not human understandable (e.g., pixels of an image), interpretable ML works explored eliciting human-interpretable \textit{concepts} in the domain (e.g., an object in the image) and use the concepts to explain the model decisions~\cite{kim2018interpretability,ghorbani2019towards}. Elicited domain concepts have also been used to create sub-groups for labeled data to enable ``structure labeling'', which could lower the re-labeling burden when a target class changes~\cite{kulesza2019structured}. We further envision elicited domain concepts could help data scientists head start their model building, as revealed in our preliminary interview. 
%\VL{here I attempted to discuss concept, please check and edit. Perhaps later when you introduce concept in Ziva, you can make the definition consistent and refer back to the citations.} 
By facilitating knowledge sharing from domain experts, we also hope to inspire novel algorithmic work that could leverage such a knowledge repository.

%\paragraph*{Concept-based learning}

%Feature engineering is a critical step to improve the accuracy of the model. Spotting a important and adjusting a weight for each feature is time-consuming and challenging~\cite{wu2019local}. 
%Beyond features, researchers proposed concepts, a unit that is sensible to a human. 
%Prior work uses concepts for model interpretability~\cite{kim2017interpretability,ghorbani2019towards} and capturing important features~\cite{brooks2015featureinsight}. %and iteratively enhancing a model (S. Amershi ‘10).
% concept-based learning and rule
% ~\cite{zweben1992learning}
% realizes concept-based learning by letting users generate concepts and enforces those concepts through a rule-authoring interface. By providing rules, users are confirming concepts and conditions that constitute the concepts. 
%CueFlik proposes interactive concept-based learning in the image domain~\cite{fogarty2008cueflik}.

%\paragraph*{Labeling tool} doccano~\cite{doccano}

%\paragraph*{Topic modeling}
%\SP{Todo for Marina}
%\MD{Incorporating domain knowledge into topic modeling is usually done as priors on topic probability distributions \cite{Andrzejewski09} 

%There's literally a background couple paragraphs on topic modeling with HIL in an IUI 2020 paper that we can take from
%\cite{Smith:Kumar:Boyd-Graber:Seppi:Findlater-2020} (\url{
% http://users.umiacs.umd.edu/~jbg//docs/2020_iui_control.pdf})}

\subsection{Technologies for knowledge sharing}

%\info{Vera: This sub-section is about tooling that facilitates knowledge sharing not necessarily specific to ML work, e.g. CSCW work on knowledge repository tools for enterprise. Building ML with domain experts' *direct* input go to the previous sub-section (e.g., interactive topic modeling, interactive ML, labeling tools)  }

Ziva is also motivated by prior work on technologies that facilitate knowledge sharing in enterprise and organizations. Knowledge sharing has been long studied in the Computer Supported Collaborative Work (CSCW) community \cite{ackerman2013sharing, schreiber2000knowledge}.
Ackerman et al. summarized \cite{ackerman2013sharing} two generations of research in this area, where the first generation focuses on the repository models to elicit knowledge as information artifacts to enable sharing, storing and re-using, while the second generation centers around expertise location and communication. Knowledge repository tools elicit various formal and informal information including manuals, standard procedures, best practices, common questions, and so on. For example, Goldberg et al. studied collaborative tagging and filtering mechanisms for workers to construct a knowledge repository \cite{goldberg1992active}; Answer Garden is a system to build a growing information repository through people asking and answering questions \cite{ackerman1998augmenting}; Terveen et al. designed a memory framework for large-scale software engineering where groups collectively build a shared memory \cite{terveen1995living}; Nam and Ackerman studied methods for elicitation of informal information into more organized forms \cite{nam2007arkose}. %\cite{schreiber2000knowledge}

% Studies have warned that knowledge sharing and repository tools often fail in practice~\cite{zhou2011cpoe,hoffmann2019cyber,yang2019knowledge}, if the design fails to take into account the organizational practices and social dynamics, including what benefits and demands these technologies bring for both the knowledge providers and the knowledge consumers~\cite{ackerman2013sharing,grudin1988cscw}.  For example, Pipek and Volker conducted a case study deploying the Answer Garden system for maintenance engineering \cite{pipek2003pruning}, where they found that the system did not work well because of the division of labor and organizational micro-politics.  Lei et al. identified many challenges in expertise sharing in inter-organizational collaboration in crisis management \cite{ley2014information} and designed an information repository system accordingly to foster collaboration.

Knowledge sharing in ML projects poses unique challenges~\cite{amershi2019software, cai2019software} to make the knowledge transferrable into ML specifications. The challenges are amplified in sophisticated domains. For example, for a medical ML model, a clinician may have to help data scientists understand complex drug information.  We inform the design of Ziva both by prior work on involving domain experts in data science projects and model development, and a preliminary interview study to understand how data scientists learn from domain experts. Meanwhile, studies have warned that knowledge sharing and repository tools often fail in practice~\cite{zhou2011cpoe,hoffmann2019cyber,yang2019knowledge}, if the design fails to take into account the social dynamics, including what benefits and demands these technologies bring for both the knowledge providers and the knowledge consumers~\cite{ackerman2013sharing,grudin1988cscw}.  Thereby  we evaluate Ziva by involving both the knowledge \textbf{consumers}--data scientists, and the knowledge \textbf{providers}--domain experts. 

%Zhou et al. looked at clinicians' workarounds to understand hospitcal specific knowledge and practices \cite{zhou2011cpoe}. Pipek and Volker conducted a case study deploying the Answer Garden approach into maintenance engineering \cite{pipek2003pruning}, where they found that the Answer Garden approach did not work well because of the division of labor and organizational micro-politics; Hoffmann et al. investigated integrating cyber-physical systems to support knowledge sharing in manufacturing context\cite{hoffmann2019cyber}; Yang et al. focused on the social interaction of knowledge transfer at work (e.g., whom to ask questions) \cite{yang2019knowledge}.
%Our work focus on tackling the technical challenges for experts to elicit key concepts in domain-specific areas.

%\AW{add a sentence on why it is challenge to elicit key concepts through traditional repository approaches}

\section{Preliminary interview}

To understand how data scientists grasp a domain, we conducted semi-structured interviews with four data scientists working on NLP models (2 females, 2 males). 
Each interview was 45 minutes long and driven by a script that covers questions related to their recent projects collaborating with domain experts and their typical interactions with domain experts. 
We recruited participants via posting on slack channels of an international technology company. Each interviewee was compensated \$15 for their time. 
We summarized our interviewees' projects and challenges in Table~\ref{tbl:prelim_interviewee}. As a result, we identified the current practices of learning from domain experts and design requirements for our tool. 
% We conducted axial coding of the interview transcripts with an open-coding protocol. The following themes emerged.

\begin{table*}
 \caption{Interviewees information}
  \begin{tabular}{p{2.5cm}|p{5.5cm}|p{6cm}}
  \toprule
  \textbf{Pn (domain)}&\textbf{Model (reasons of choosing the model)}&\textbf{Methods of knowledge exchange}\\
        \midrule
        P1 (Legal, law)  & Rule-based (transparency, few labels)  & Instance perturbation %\texttt{data.peek(3)}
        \\
        \midrule
        P2      
        
        (Disaster recovery) & Supervised neural net (sufficient labelers) & Education session of domain overview domain experts labeling%\texttt{data['result'], data['confidence']}
        \\
        \midrule
        P2 (") & Rule-based (transparency, few labels) & Domain experts think aloud labeling data %\texttt{feature X in data['content']}
        \\
        \midrule
        P3 (Customer categorization) & Random forest (transparency)& Pair-authoring (Go over analysis together with domain experts ~\cite{wang2019data})
        \\
        \midrule
        P4 (Sports) & AutoML (time) & Brute-force model building
        \\
      \bottomrule
    \end{tabular}
   
    \label{tbl:prelim_interviewee}
    \vspace*{-5mm}
\end{table*}

%\info{Vera: say something , perhaps by the end, about "knowledge sharing" is not just about labeling data but also other parts in ML development, sometimes open-ended learning -> thus siva is not a labeling tool but learning tool for DS. The design requirements for domain experts is efficiency, low workload (thus structured elicitation methods for desired formats), and asynchronous exchange with data scientists (since limited availability). For data scientists, we (preliminarily) identify concepts and rationale are useful knowledge to learn in NLP domains, so the first version of Ziva will focus on them (I positioned in the intro as a "first version" and the later interview study will inform future development. this might help defending against reviewers questioning some kind of domain knowledge is left out)}
% \info{Soya: I added the part based on Vera's comment in the ``Summary'' paragraph, please check}
\subsection{Limited time and limited best practices}

All of our data scientist interviewees indicated they often need domain experts' help and feedback. 
% \textit{``I was given enough knowledge to do that, but there were, of course, many documents which were complicated to assign one category. So at that time, those SMEs come into picture who have 10 years of experience in the field.''} 
However, domain experts are busy and have little time to spare. One said:
\textit{``The first issue is getting hold of their time... I think hardly I was getting one day a week, you can say one hour a week, not even an entire day.''} %\textit{``SMEs have a lot on their plates, and this is just one part of their job where they can assist me in building a tool like this. So I think the first issue is getting hold of their time. So initially it was difficult, I think hardly I was getting one day a week, you can say one hour a week, not even an entire day.''}
So data scientists try to extract as much knowledge as they can in the limited time they have. They have to spend significant time preparing for these discussions.  For example, they often manually curated examples such as mis-classified instances and instances that contain the unfamiliar keywords to ground the discussion during the meetings with domain experts. 
%\textit{``I think I was myself planning to use LIME or something like that to make explanations clear, because clients don't understand neural network. We have an LSTM employed in that. So I don't know reason why they would trust the judgment coming from it. So I think in that case, a good thing would be that, okay, I'm classifying this as action because, okay, these is probably the word which has a lot of page to it, or the context. So whatever is the reason for it to assign it to a certain category, I think I would want it to highlight it.''}
%It is a high learning curve for data scientists to learn a domain, even more if the domain is specialized. 
% Once data scientists equipped the basic knowledge of the domain, they established communication methods.
Even though there is no standard way to extract domain knowledge across different domains, but through \textit{mutual effort} they find what works best for a project. We identified the following approaches to learn domain knowledge from domain experts:  

\textbf{Example-driven conversation:}
The first fold of approaches is domain knowledge sharing based on examples. By inquiring about how and why domain experts would label or make decisions for these examples, data scientists learn \emph{rationales} of how the model should behave for the instances.
There are three tactics mentioned by our interviewees. 
P2 observed domain experts during labeling to learn the domain experts' thought process: \textit{``They would go line by line in front of me so that I can also see what their brain is looking at classifying them.''} 
P1 initially took P2's approach, but due to the complexity of the law domain, explaining rationale required extensive background knowledge. Oftentimes, it is unclear to data scientists how to connect the explanations provided by the domain experts to model specifications. %Over time, the team evolved a strategy for how to give concise feedback that is useful to data scientists. 
% The strategy followed focused on identifying the most important tokens in the example in the model's decision making. 
P1 used a strategy called \emph{instance perturbation} --
for a given instance, the domain experts were asked to minimally change the instance until the model changes the label and discuss the reasons. 
With this, data scientists were able to narrow in on the parts of the instance that should be the most important to the model's decision. 
Instead of aiming to build a perfect model right off the bat, P4 deployed their model first and incrementally improved the model upon domain experts' request. Whenever domain experts encountered mis-classified results, they shared the instances with data scientists and discussed why they were mis-classified.  

\textbf{General background knowledge acquisition:}
Concepts are key units of information for a given domain, such as notions, entities, components or properties. A set of domain concepts can be seen as a \textit{taxonomy}. Understanding them could help data scientists make sense of the domain.  Participants reported approaches to learn concepts in an unfamiliar domain. 
P2 and P3 said domain experts in one of their projects offered a \emph{lecture} to explain key concepts their domain. For P2, domain experts gave an overview and touched on the basic concepts of each class. P3  ~\emph{pair-authored}~\cite{wang2019data} with domain experts to bridge concepts and a mathematical formula that encapsulates the information. With this iterative learning process, data scientists were able to kick start model building. 
P2 said \textit{``I think that was very helpful because after that, my dependency reduced a bit. I could myself assess that what category they belong to.''}
% domain experts provided the definitions and \emph{data quality} of each column. For instance, domain experts were able to provide insights about how noisy or sparse each column. 
%The other two participants, received no explanations from domain experts and resorted to search for relevant information on the internet to teach themselves.

% \DP{...to here. The reason being that although the content above is interesting, it's not really used in the paper. I think the take away is that there is no set strategy for doing this exchange of information, and it can be said in less words.}

% \VL{R1 and R2 are a bit too high-level. Can you say something more specific, e.g., for R1, asynchronous collaboration, able to re-use, efficiently with scaffolding for desirable format; For R2, what basic knowledge? if it is just what's in R3, then omit R2...  right now Ziva still feels under-motivated}\SP{R2 is motivation for concept and R3 is for rationale}
\paragraph*{\textbf{Summary:}} From our interview, we derived several design requirements to design Ziva. We found that the usage of domain knowledge is not only limited to labeling but also other parts in ML development, sometimes open-ended learning. Thus, interfaces of Ziva ought to facilitate domain-knowledge learning of data scientists in general throughout the development workflow. More specifically, we found that the tool should scaffold domain experts to efficiently elicit domain knowledge within short amount of time (\textbf{R1}). Next, a tool should help data scientists to extract basic domain concepts (\textbf{R2}). Lastly, data scientists indicated that they often learn from domain experts' rationale, especially how they justify a decision or label. Hence, the tool needs to facilitate label justification sharing (\textbf{R3}).

\section{Ziva: Interface for Eliciting Domain Knowledge}

\begin{figure}
  \centering
      \includegraphics[width=1.05\columnwidth]{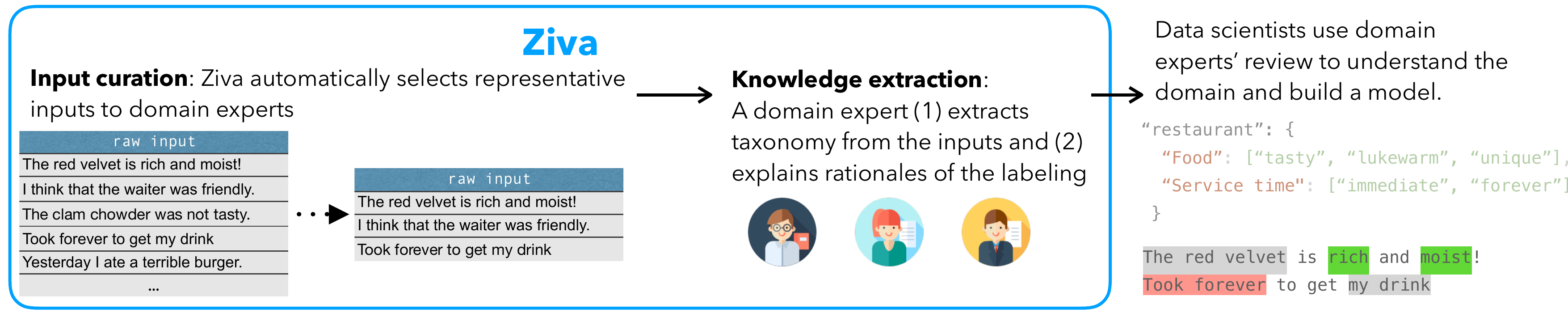}
  \caption{\rev{To facilitate domain knowledge sharing, Ziva presents representative instances and to interfaces to review the instances to domain experts, then which will be used by data scientists. }}
  \label{exit-survey}
  \vspace{-7mm}
\end{figure}

This section introduces the interface of Ziva. Ziva provides features for domain experts to create domain concepts and elicit justification from representative instances that are automatically curated by Ziva. 
We discuss Ziva's different components and how they meet the design requirements in detail.

\subsection{Representative sampling for instances creation}
\label{subsection:representative_sampling}

As highlighted in the our formative interview, domain experts have limited time for labeling or sharing domain knowledge (R1). Hence, it is important to ask them to review only a few of instances and the sample can cover most concepts in the domain. Ziva extracts such a representative sample of $m$ instances from a large training set of $N$ text instances by the simple method of transforming the original text into 'tf-idf' space, clustering the result using an algorithm such as k-means (setting $k=m$), and, for each cluster, returning the text instance closest to the cluster center. This method is not deterministic, but provides a reasonable set of representative instances, for cases where $m<<N$.

\subsection{Concept creation}

\begin{figure}
  \centering
      \includegraphics[width=1.05\columnwidth]{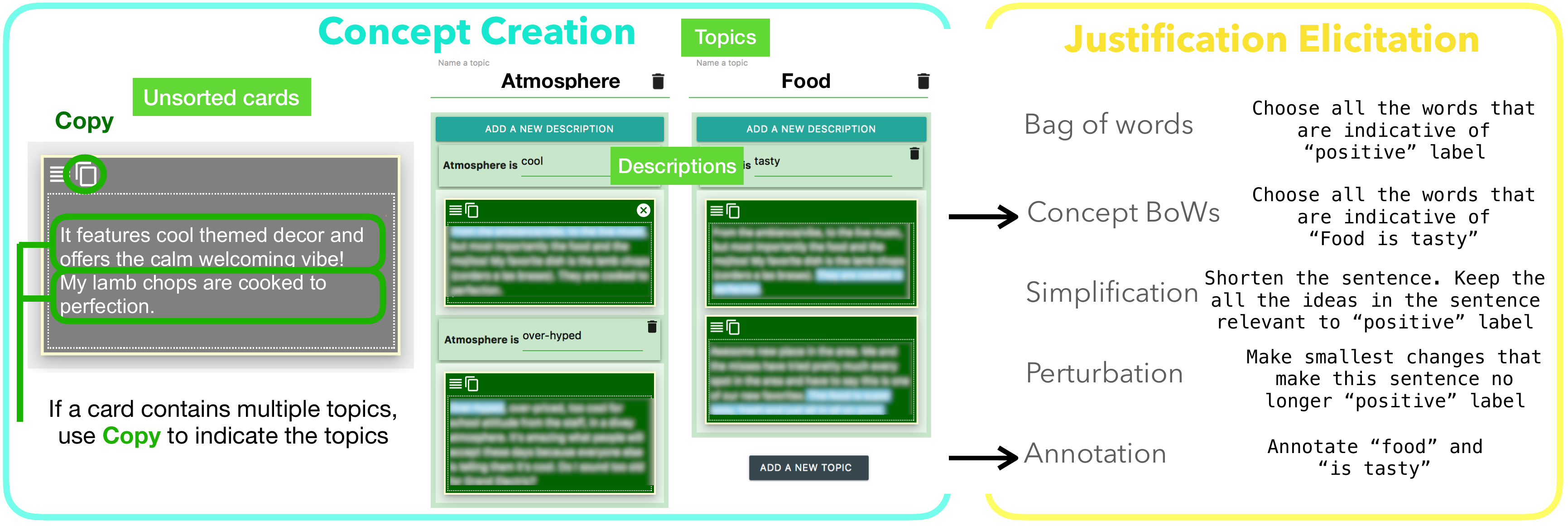}
  \caption{Ziva interface: domain experts first extract domain knowledge with curated instances. Then they review each instance one by one using one of justification-elicitation interfaces.}
  \label{fig:interface}
  \vspace{-5mm}
\end{figure}

Creating a taxonomy is an effective way of organizing information~\cite{laniado2007using,chilton2013cascade}. Ziva provides an interface where SMEs can extract domain concepts (R2). Users are asked to categorize each example instance, presented as a card, via a card-sorting activity. Users first group cards by topic (general concepts of the domain such as atmosphere, food, service, price). Cards in each topic are then further divided cards into descriptions referencing specific attributes for a topic (e.g., cool, tasty, kind, high). The interface (Figure \ref{fig:interface}) was implemented as a drag-and-drop UI using LMDD \cite{lmdd}.

\subsection{Justification-elicitation interface}

Once a domain expert finishes the concept extraction, they review each instance using one of elicitation interfaces, which ask the domain expert to justify an instance's label (this information is then intended for consumption by data scientists (R3)). We used Materialize to implement the  justification elicitation conditions.

The \name{justification elicitation} interfaces were designed through an iterative process of paper prototyping, starting with initial designs inspired by our preliminary interviews. As we conducted paper prototyping, we examined if (1) the answers from different participants were consistent and (2) the information from participants' answers were useful to data scientists. 
%As a result, we came up with 5 conditions as mentioned above and the design rationale for each \name{justification elicitation} are as follows:
We now describe the five different \name{justification elicitation} methods that we created and evaluated, and highlight the design rationale where appropriate.

%\paragraph*{\just{Bag of words}}

\textbf{Bag of words.} This base condition reflects the most common current approach. Given an instance and a label (e.g., positive, negative), the domain experts are asked to highlight the text snippets that justify the label assignment.

%\paragraph*{\just{Instance perturbation}}

\textbf{Instance perturbation. }Inspired by one of our data scientists in the formative study, this condition asks a domain expert to \emph{perturb} (edit) a part of the instance such that the assigned label is no longer justifiable by the resulting text. For example, in the restaurant domain, \textit{``our server was kind''}, can be modified to no longer convey a positive sentiment  by either negating an aspect (e.g., \textit{``our server was not kind''}) or altering it (e.g., \textit{``our server was rude''}). 

This strategy is also inspired by the research area of generating natural language adversarial examples \cite{alzantot-etal-2018-generating}. Such approaches algorithmically alter training examples to create similar adversarial examples  that fool well-trained NLP models. In our scenario, the domain expert is seeking to alter training examples in order to point out the most salient characteristics to the data scientist; the latter learns from this information, combining it with syntactic and semantic analysis of the original and perturbed instances.

%\paragraph*{\just{Instance simplification}}

\textbf{Instance simplification.} This condition asks domain experts to shorten an instance as much as possible, leaving only text that justifies the assigned label of the original instance. 
%Domain experts can choose to not shorten any part if they think the entire part are important. 
For example, \textit{``That's right. The red velvet cake... ohhhh.. it was rich and moist''}, can be simplified to \textit{``The cake was rich and moist''}, as the rest of the content does not convey any sentiment, and can therefore be judged irrelevant to the sentiment analysis task.

This condition is inspired by the plethora of methods for sentence simplification used in extractive text summarization \cite{Vale20}. In particular, the domain expert is performing \textit{sentence reduction} as in \cite{Jing_2000}. The output can be considered to be a concise summary of the original instance, keeping only that content which is directly relevant to the sentiment analysis task. The result for the data scientist is clean, compact, and fully relevant high quality training examples.

%\paragraph*{\just{Concept bag of words}}

\textbf{Concept bag of words. }This condition incorporates the concept extracted in the prior step. Similar to the \just{Bag of words} condition, domain experts are asked to highlight relevant text within each instance to justify the assigned label; however, each highlight must be grouped into one of the concepts.
%Similar to the \just{Bag of words} condition, we ask domain experts to highlight relevant text for each instance. Then, they were additionally asked to select the words in that text that are related to concepts they built in the \just{Concept creation} interface. By doing so, domain experts embed concepts for each instance. 
If, during \name{Concept creation}, the domain expert copied a card to assign multiple topics and descriptions, then the interface prompts multiple times to highlight relevant text for each one. For example, if they classified the instance, \textit{``That's right. The red velvet cake... ohhhh.. it was rich and moist''}, into the concept \textit{``food is tasty''}, they can select \textit{rich}, \textit{moist} and \textit{cake} as being indicative words for that concept.

%\paragraph*{\just{Concept annotation}}

\textbf{Concept annotation. }This condition is similar to the above \just{Concept bag of words} condition. However, when annotating the instance text, domain experts are directed to distinguish between words relevant to the topic and words relevant to the description. Given the above sample instance, the domain expert would need to indicate which part of the sentence applies to \textit{food} (e.g., \textit{cake})  and which to \textit{tasty} (e.g., \textit{rich and moist}). 
Both this and the previous concept condition are motivated by the well-established knowledge that a variety of NLP tasks, such as relation extraction, question answering, clustering and text generation can benefit from tapping into the the conceptual relationship present in the hierarchies of human knowledge \cite{zhang-etal-2016-learning}. Learning taxonomies from text corpora is a significant NLP research direction, especially for long-tailed and domain-specific knowledge acquisition\rev{~\cite{wang-etal-2017-short}}. 

In the rest of the paper, we present a case study to evaluate the utility of the Ziva interface in two parts. In Section~\ref{sec:domain_expert}, we conduct a lab experiment and a crowd experiment in which participants acted as domain experts using Ziva. We choose the domain of  restaurant reviews (Yelp Open Dataset\cite{yelp}) and the NLP task of sentiment analysis, as being extremely familiar and easy enough to understand for most people to qualify as domain experts. In Section~\ref{sec:data_sientist}, we conduct \rev{an interview} study with data scientists to evaluate the utility of domain knowledge collected in the above experiments.  We instruct the data scientists to assume no previous knowledge of the domain, so we could use the elicited knowledge about restaurant reviewing as proxy to understand how Ziva could help them build NLP models.

%as a target domain since within the domain, is easy enough to understand with common sense which makes participants to act as domain experts using our tool. There are several components in the domain, this led us to proxy what are the challenges in building an ML model in a sophisticated domain. 

\section{Evaluation on domain experts' experience}

\label{sec:domain_expert}
%To assess the elicitation methodologies, we invited participants to use our Ziva interface. To simulate domain experts, we recruited participants who are not knowledgeable in ML. 
We recruited participants to act as domain experts of restaurant reviewing to use Ziva. In a lab study (N=12), we compared participants' task completion and experience with all concept and justification elicitation methods, and gathered their qualitative feedback. To allow quantitatively compare the results of different justification elicitation methods, we conducted a follow-up crowd experiment (N=88).

%\DP{It's unclear to me how this study relates to the one in the section above. I think it would help the story if the purpose of it as it relates to others would be made more explicit. I think what's missing is that it's unclear how the rankings and takeaways from the paragraphs above relate to this.}
% \begin{table}[t!]
%     \caption{Demographics of participants}
%     \label{tab:demographic}
%     \begin{tabular}{c c c}
%         \toprule
%         \textbf{Gender} & \textbf{Job Role} & \textbf{Work Experience in years}\\
%         \midrule
%         M & Project Manager & 10 years \\
%         M & Software Engineer & 20 years \\
%         W & Researcher & 2 years \\
%         W & Software Engineer & 2 years \\
%         M & Trained Professional & 8 years \\
%         M & Graduate Student & 1 year \\
%         W & Designer & 1 year \\
%         M & Software Engineer & 17 years \\
%         W & Skilled laborer & 28 years \\
%         M & Software Engineer & 16 years \\
%         M & Researcher & 38 years \\
%         W & Software Engineer & 22 years \\
%         \bottomrule
%     \end{tabular}
% \end{table}
\subsection{Lab study}
\textbf{Study protocol}:
% \subsection{Dataset}
%Using our representative-instance curation method, we extracted 10 reviews.
%The reviews contain 3,057 letters and 642 tokens. 
To avoid noisiness in labeling, we pre-labeled the set of yelp reviews instance so we could focus on comparing the elicitation methods.  We created binary labels based on ground-truth ratings: if the number of stars is 1 or 2 for a review, we labeled it as negative, 4 or 5 as positive~\cite{zhangCharacterlevelConvolutionalNetworks2015}.
We then took a random balanced sample of 10,000 instances. 8,000 were used as a (balanced) 'training set', from which we extracted ten representative instances to use in the study (see Section \ref{subsection:representative_sampling}.) We set the other 2,000 (balanced) instances to use as a test set for analyzing the performance of models built from the study output (see Section \ref{subsections:learning_outcome}).

% \subsection{Participants}

We recruited participants (5 female, 7 male), who self-report little or no knowledge in ML via posting on slack channels of an international technology company. Participants are designers, graduate students, researchers, trained professionals, skilled laborers, software engineers and project managers. % (Table \ref{tab:demographic}).
To compensate their time, we ran a \$30 raffle.

Participants were given introduction to the project and a tutorial of the Ziva interface. They were also given a practice task in a different domain, i.e., clothing. For the concept extraction task, all participants used the same interface. For the justification interface, we randomly assigned each participant one treatment from the elicitation methods without concepts (bag of words, label perturbation and simplification), and one from those with concepts (concept bag of words and concept annotation). Thus, each participant experienced two elicitation interfaces and reviewed 5 instances each. After each interface, participants were asked to fill out the NASA TLX form \cite{hart1988development} to evaluate their subjective workload and share their feedback. The entire session lasted about for up to one hour. 

\textbf{Task Results}: One participant could not complete the second \name{justification} interface. We reported the summary of \name{concepts} generated by participants, as well as quantitative and qualitative experience using \name{justification} methods.   
\paragraph*{Concept creation}

Participants took 879.7 seconds on average ($\sigma$=385.4). They created 3.92 topics on average ($\sigma$=1.11). Everyone included \textit{Food quality} and \textit{Customer service} in their topics. To assess taxonomy from each domain expert, we examined the consistency between domain experts and coverage of the restaurant domain.   

\begin{itemize}[noitemsep,topsep=0pt]
\item Consistency between domain experts: The union of all topics across all participants includes following 10 topics: ambiance, cuisine, food quality, customer service, additional service, complaint, speciality, reservation, location, and price. For each topic, we rated whether each taxonomy intersects with the topic or not. 
Thus, the inter-rater reliability (IRR) across all domain experts was 58\% using Fleiss’ $\kappa$. 
\item Coverage of the domain: We selected 3 additional instances which were not shown to the participants. We used our curation method to pick another set of representative instances.
We then inspected how many instances can be categorized using each taxonomy resulting in a coverage of 69\% (25 out of 36 instances).  
\end{itemize}

\begin{table}[]
\caption{Average task completion time (standard deviation) of lab study participants}
\begin{tabular}{c|c|c|c|c}
\textbf{Bag of words} & \textbf{Simplification} & \textbf{Perturbation} & \textbf{Concept bag of words} & \textbf{Concept annotation} \\ \hline
39.2 s (20.7)             & 106.6 s (86.8)             & 107.2 s (48.0)             & 36.8 s (14.0) & 81.9 s (40.5)    \\ %\hline
\end{tabular}
\label{labstudy-time}
\vspace{-7mm}
\end{table}

\paragraph*{Justification elicitation}

The average task completion time in each condition is summarized in Table~\ref{labstudy-time}. Since each participant was assigned two out of five justification elicitation, there were only a few data points per elicitation technique (3 to 5 per technique). To further investigate in a larger population, we deployed Ziva on a crowd platform described in the following section. Plots of the post-questionnaire results are attached in supplemental materials.

Most participants found  the \just{bag of words} condition easy to complete. One participant said: \textit{``This was easy because a lot of words were clearly positive or negative, such as "terrible" or "delicious"''}. However, some considered it tricky to identify words that are indicative of the overall sentiment. For example, one participant said, \textit{``this can be just descriptive without any positive or negative feelings without the context. So it's difficult to isolate the context out of the words.''}

For the \just{simplification} task, participants indicated the task was straightforward. Participants said \textit{``easy as it had eliminated redundant and unnecessary words''} and \textit{``quite easy and intuitive, paraphrasing keeping the original intent is what I usually do as part of minutes of meetings''}. One participant said sometimes the task became hard because some instances could not  be obviously shortened and instead need to be entirely rewritten.

%\textit{``Often the sentence as presented would not be how I would convey the topic. I was not as clear if I could rewrite vs. trim with minor edit''}.

Participants said \just{perturbation} is also straightforward but it required them to understand the entire instance thoroughly. One participant commented, \textit{``It was kind of hard because I don't know some of the words''}. Another participant suggested that if the interface suggested antonyms, it would be easier to finish the task. 

With \just{concept bag of words}, participants said it allowed subjective and nuanced elicitation, as they could pick words associated with a \name{concept} without judging their sentiment.  %\DP{Here would be a good place to explicitly state \textit{why} it enabled nuance in addition to the example that follows.} 
However, it led to more varied results among participants. For example, for the concept \textit{Food is tasty}, and the instance \textit{Ohhhh... The red velvet cake is rich and moist}, most participants selected \texttt{rich} and \texttt{moist}. One participant said \textit{``Even \texttt{red velvet cake} could be the indicative words if you personally like the cake''}. Others said \textit{``maybe \texttt{ohhhh} part can be included''} and \textit{``\texttt{moist} doesn't necessarily mean delicious''}.

For the \just{concept annotation} task, participants said it is straightforward to choose words directly mapped to each token.
%, one said \textit{``Easy as the comment had straight forward message''}. \DP{The previous quote doesn't stand on its own. You'll have to add some context} 
On the other hand, it complicated the articulation to have to label in such fine granularity. One participant commented, \textit{``slightly tedious as it required me to comprehend on how best to label the words accordingly''}.

% \begin{figure}
%   \centering
%       \includegraphics[width=.5\columnwidth]{figs/labstudy-time.pdf}
%   \caption{Average task completion time. Error bars indicate standard deviations (lab study participants, n=12)}
%   \label{labstudy-time}
%   \vspace{-6mm}
% \end{figure}

\subsection{Crowd Experiments}
\label{subsection:crowd_experiments}

To assess different \name{justification} methods on larger population, we deployed the Ziva interface on a crowd platform.  

% \subsection{Dataset}
\textbf{Study Protocol}:
Using our curation method, we extracted 10 reviews from the datasets used in the lab study. We pre-populated a taxonomy. In order to provide a representative sampled \name{concept}, we recruited 5 volunteers and asked them to extract concepts of the restaurant domain using the \name{concept extraction} interface and two of the authors aggregated the taxonomy. 
The resulting taxonomy is attached in supplemental materials.

We installed 5 test questions  for each condition with ground-truth created by the authors. If a crowd worker did not pass more than half of test questions, they can not continue to the Human Intelligence Task (HIT). Each worker was given one of the five justification elicitation interfaces, and reviewed 10 instances.

We recruited our study participants from Appen~\cite{appen}. We compensated them with \$0.5 per HIT, they are rewarded \$2.5 in addition for test questions. 
From the lab study, we observed each HIT took less than 2 minutes on average, which makes hourly wage of \$15. After the tasks, we asked them to fill out the same NASA TLX form to report on their subjective workload. Participants were rewarded additional \$3 for the survey.   A total of 88 crowd workers completed our study resulting in 857 instances with elicitated data.
% bag of words 17 simplification 17 perturb 17 conceptbow 18 concept annotation 19

\textbf{Result}:
We analyzed participants' survey responses using an one-way Kruskal–Wallis ANOVA as summarized in Table~\ref{crowd-anova}. There was marginal difference in self-reported success of task accomplishment and significant difference in stress level across justification elicitation methods. 

As a post-hoc analysis, we ran a one-tailed Mann-Whitney U Test. The result revealed that participants completed the tasks using \just{bag of words} perceived higher success in accomplishing the tasks than participants with \just{simplification} (U=76.5, z=2.31; p=.01) and \just{concept annotation} (U=97.5, z=2.02; p=.02). \just{Concept bag of words} users also perceived higher success than \just{simplification} (U=75.5, z=2.34; p=.009) and \just{concept annotation} users (U=103, z=1.85; p=.03). 

As for the stress level, \just{bag of words} users reported significantly higher stress than \just{perturbation} (U=55, z=2.90; p=.002), \just{concept bag of words} (U=84.5, z=-2.24; p=.01), and \just{concept annotation} users (U=81.5, z=-2.34; p=.01). 

\begin{table}[]
\caption{Crowd experiment Likert result. H statistics (p-value) in significance level 0.05}
\begin{tabular}{|c|c|c|c|}
\hline
\textbf{Mentally demanding} & \textbf{Successfully accomplishing} & \textbf{Hard to accomplish} & \textbf{Insecure, Stressed} \\ \hline
2.0825 (.72059)             & 8.7959 (\textbf{.06641})             & 8.0609 (.08937)             & 9.9411 (\textbf{.04143})     \\ \hline
\end{tabular}
\label{crowd-anova}
\end{table}

\begin{figure}
  \centering
      \includegraphics[width=1.05\columnwidth]{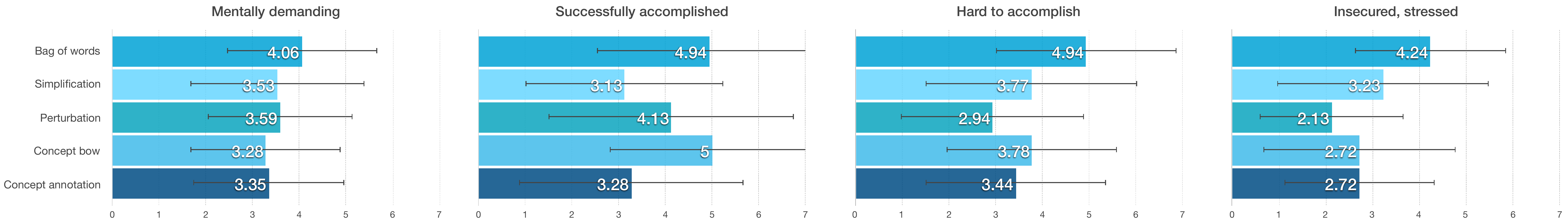}
  \caption{Post-question responses in NASA TLX (1- Very low, 7- Very high) (Crowd experiment participants, n=88)}
  \label{exit-survey}
  \vspace{-5mm}
\end{figure}

\section{Data scientists \rev{interview study}}
\label{sec:data_sientist}
To investigate what and how domain knowledge extracted from Ziva helps data scientists, we conducted an interview study with data scientists. We showed them \name{concept} and different \name{justification} results extracted by domain experts and asked them how they could use them in their ML development workflow. 

% \DP{Agree with Vera's proposed strucuture here}

% \info{Vera: this is the most important section of your paper. It should ideally both validate Ziva (different usage for building ML models, pros and cons of different elicitation methods), and discuss some future opportunities or design requirements for Ziva. Probably one subsection about different usage of domain knowledge (I don't think you have to discuss current practice in separation), one section about ranking and feedback for each technique, and one section summarize some design requirements (e.g., scalability. here you can mention current practice, but overall your goal is to talk about design requirements, which you might not be able to uncover from the preliminary study). If you see in the data about other types of knowledge or elicitation methods useful, could be helpful to mention here too}

\textbf{Study Protocol:}
Participants were given introduction to the project, prompts shown to domain experts and corresponding outputs of each part of the interface.
Each interview was 1 hour long and driven by a questionnaire that posed questions related to compare domain knowledge extracted by domain experts using Ziva to their current practice. Finally, they were asked to rank  usefulness of \name{justification} interface to their workflow.

% \subsection{Participants}

We recruited 7 data scientists who have between 4 and 20 years of experience building models with domain experts in sophisticated domains, using the slack channels of an international technology company and word-of-mouth. 
%The participants' experience  ranges from 4 to 20 years.

\textbf{Results:}
% \subsubsection{Ranking of \name{justification} techniques}
% We asked interviewees to rank 5 \name{justification} techniques of Ziva based on the usefulness of each technique in their own workflow. 
The full ranking of each technique is reported in supplemental materials. We re-ranked the scores on a linear scale, with a data scientist's favorite at 5 points, the second-most favorite at 4 and so on. If two techniques were tied for N-th rank, we averaged the scores for the both of techniques (e.g., if two techniques are 4th, they are given 1.5) As a result, the \just{concept annotation} technique scored the most (30), then \just{concept bag of words} and \just{perturbation} (22.5), \just{simplification} (17.5), and \just{bag of words} (12.5).
% Data scientists discussed their preference on each \name{justification} technique and various use cases of the techniques in their workflow. 
%Here, we focus on the \emph{metrics} that data scientists to rank the \name{justification} techniques. 
Data scientists had several reasons why they prefer one \name{justification} technique to another and various applications for different techniques. Through the metrics, we were able to identify the design requirements and important factors of domain-knowledge learning.

% \subsubsection{Current practice}

% Data scientists shared their current practice of domain-knowledge learning. As a result, two of data scientists used \textit{input perturbation}, other two received \textit{education session} from domain experts, and one took \textit{brute-force model building}. 

\paragraph*{Standardized protocols}

As revealed in our preliminary interview and previous work~\cite{mao2019data}, there is no set protocol of communications or \emph{common ground} between two parties, and expressed a need for a \emph{protocol of communication with domain experts}. 
The steep learning curve of a specialized domain and lack of guidance in how to extract domain knowledge  exacerbates the collaboration with domain experts. %Especially, a bottleneck of on-boarding an ML project is problem statement. From users' request, data scientists had to derive ML specifications: \textit{``They [domain experts] did not really have a well-defined problem in mind''}. In order to set a sensible goal, it is necessary to know the domain. %\textit{``So from my perspective it's actually like coming up with ways that allows me to understand the domain better the outcome.''}
% One interviewee said,
% \textit{``We wanted to start simple. So we wanted to understand generally about [the domain]. But ideally we also want to again define - like just make more and more questions from me. But then they're saying their end goal.''} 
Three of interviewees said that having such a \name{concept} and examples upfront provided by domain experts has helped them to build a model in prior projects.  One said, \textit{``They describe what are the component information and examples. It was not very difficult to understand after reading the documents.''} 
In light of this, interviewees preferred \name{justification} techniques to inform them about the domain. For example, interviewees found \just{concept annotation} helpful because it is tightly connected with the \name{concept}, hence they can learn from examples how different components of the domain is expressed in the instances. \just{Simplification} is also helpful, as it is a simpler version of instances without rhetoric.

% \textit{``I try to ask the same question over and over and over but in different ways.''}

% \textit{``It would actually help me to understand the domain as well. I have to get help from the domain expert because I got it and I don't understand why the examples look extremely identical but they have different labels.''}

% \textit{Explainability} 

One interviewee suggested to use \name{justification} techniques to \emph{explain} model decisions. They said, \textit{``I work on active learning ML a lot. So I work with users. And so far all the interactions I expect for the user, fairly simple, either binary feedback -- correct or incorrect. I have any incorporated explanation of when the user provide feedback. What's the explanation behind this feedback? I think that that would be very useful to generate some explanation or learn how to generate explanation.''}
While model explanation is not the intended usage of the Ziva's \name{justification} techniques, the data scientist found the techniques helpful for debugging a model. 

\paragraph*{Scalability of domain knowledge}

Interviewees were also interested in how they would \emph{scale} the Ziva output. Since they only received only 10 labeled instances and justification, it was too small for data scientists to train a model. 

% \emph{direct} and \emph{indirect model building}.

% \paragraph*{Bootstrapping labels}

One application of Ziva output mentioned by data scientists is to label more instances by generalizing \name{concept} and \name{justification}, so called \emph{weak-supervised learning}~\cite{ratner2019weak}. One interviewee said, \textit{``They are trying to give me guidance on how to propagate the labels. So one, the concept is going to be able to give me some notion on how to bucket my data, right, like, just in an unsupervised fashion.''} 
% In light of this, interviewees said that \just{concept annotation} was helpful since it tightly connects \name{concept} and \name{justification}.  
% One said, \textit{``This combines concepts of the taxonomy with the concept annotation. So it, like, combines both. I can take those and rules, convert them to rules and propagate the labels across my unannotated data set.''}
% \SP{quote for using perturb for labeling} 

% \paragraph*{Feature engineering}

Interviewees stressed the importance of domain knowledge in \emph{feature engineering}. During meetings with domain experts, they focus on identifying features for their model: \textit{``I immediately start looking at what are the different features or abstractions of features that seem to be important to the domain expert.''} However, data scientists expressed the difficulty of feature ideation in building models in a specialized domain. Repeat meetings were required to go over many instances together in hope of covering the complete set of features. 3 of interviewees said they would use Ziva output to facilitate feature engineering by using the concepts created by domain experts as features. A participant expalined: \textit{``Vector that we can convert each restaurant record into a some feature vector.''} When it comes to the best \name{justification} techniques for feature engineering, one said 
\textit{``The one with the highest resolution would be more beneficial for feature learning potentially because it allow me to generalize better''}. One data scientist suggested that they can propagate the feature across different components (e.g., food/food quality, service) of the domain expert's concepts using \emph{distributional signature}~\cite{bao2019few}. For instance, in a restaurant domain, once they identified positive-sentiment words related to food, they can find similar sentiment of words related to service using the distribution of words. 
% \just{concept annotations} are more \textit{more fine-grained features}.

% \textit{``Even train a very simple model for feature learning, just like the distributional signature''}

\paragraph*{Reduced burden on domain experts}

We also found data scientists were being mindful of domain experts' cognitive load when they generated Ziva output, because domain experts were often busy. Another reason is if the eliciting \name{justification} is difficult, data scientists would not get a reliable result. One interviewee said: 
\textit{``I would say there's also the question of what I think would be more easier for people, if it's difficult, then they're probably not going to do it very well. I wouldn't give it to them because I would think it's going to be more noisy.''} %We continue investigating on the cognitive load of domain experts when they generated domain knowledge using Ziva in the following section.

% \textit{``how do we come up with in our presentation of how to propagate the label better, right? So it's not only going to give me the resolution of food quality.
% Like if it's food quality I need to find the word taste in the sentence before I label it so I can have - there are labeling functions or rules that I can to propagate my label.''}

% \textit{``So then I started building in my mind, okay, time is an important factor. It may be difficult for me to understand, like, how can I extract this feature because it's not in the data. So then I would say maybe it's actually very important to start to figure out how to parse these sentences that could lead me to create such a taxonomy in my mind?''}

% One data scientist said: \textit{``I would start extracting, like what is it about this particular example that they've shown me that tends to be positive or negative?''} 

\paragraph*{Elicitation and learning outcomes}
\label{subsections:learning_outcome}

%In order to identify and isolate the signal of how the different label-justification elicitation methods affect the quality of downstream NLP models, we chose a very simple rule-based NLP model and a simple sentiment task. The task to identify the polarity of instances. We also chose a balanced-classes setting. Our train and test sets are balanced. In particular, our training set has 8,000 total instances, with equal positive and negative instances. The results in Table \ref{tab:rulemodel_output} is on a hold-out test set with  1,000 positive instances and 1,000 negative instances. 

%As a reminder, we only have 10 instances labeled and annotated using Ziva. Our goal is to use Ziva to learn a rule-based NLP model that propagates labels to the rest of 8,000 instances in the training dataset, and predicts the labels of  the 2,000 instances in the test dataset. This is a challenging task to any ML model. However, this is the scenario that we repeatedly see in real-world settings. 

% We now try to identify the effect of using each Ziva justification elicitation methods to build a sentiment analysis model for classifying restaurant reviews.
To demonstrate the feasibility of translating the Ziva output into useful features for model building, we constructed a real implementation. Inspired by a use case suggested by a data scientist in our study, we built 5 rule-based models for weak-supervised learning, 
%We used the results generated by our participants of crowd experiments on 10 instances, 
%Recall that the 10 analyzed instances were chosen to represent a 'training set' of 8,000 instances, leaving a balanced test set of 2,000 instances. We implement a  rule-based model, which can work with even a tiny amount of labeled data, and is fully transparent, allowing for thorough analysis.
%It is critical to understand that we are NOT seeking to create a good quality sentiment analysis model. 
%Second, and most interestingly, we want to discover how far we can get with only 10 training examples, 
mimicking a real-world cold start scenario with extremely limited labeling resources and no pre-trained model available. With such constraints, no one can expect state-of-the-art performance after a few training examples. Instead, a valuable characteristic at this early stage is intra-class consistency, demonstrating parallel improvement in precision and recall performance on various classes (here, positive and negative sentiment). This would suggest that the model is indeed learning something relevant to the entire task rather than guessing wildly, and hints at a good robustness that can be reliably improved upon with additional examples.

% Soya: I propose to move the part that I commented out to supp material Marina: yea I agree
% For each condition, the model analyzed each instance as dictated by the information provided in the condition, and detected any signals that would indicate a positive or negative sentiment label. If more positive signals were detected than negative signals, a positive label was assigned (and vice versa). If no signals, or an equal number of positive and negative signals were detected, no labels was assigned. 

Excepting the \just{bag of words} condition, %since the task is sentiment analysis, 
the models primarily focused on recognizing the semantic pattern of `Noun is Adjective'. %mimicking the process of \name{concept creation} for this task. 
Of course, this can take several forms (`food is good', `good food', `food is not bad', etc.) We built rule-based models that extend a generic semantic role labeling model \cite{Akbik2016KSRLIL} which can easily handle such variations. The generic model identifies all existing semantic roles, and the ten instances, annotated in each condition, are used to populate the dictionaries that those roles should match on (e.g., `food' and `good').
%It is important to note that we stuck to the vocabulary directly present in the instances, to present the cleanest comparisons. If we were actually seeking to build a high quality model we would of course investigate extending the training vocabulary using embeddings or other external resources. However, this is well beyond the scope of this paper; here, we seek only to show what can be accomplished with the output of Ziva. 
In general, we were careful during model construction to not make any use of additional external knowledge (e.g., we do not know that `hot wings' and `burgers' are both a type of food, if this information was not in the output of Ziva.) Below we describe the details of each elicitation method and discuss the results, which are summarized in Table \ref{tab:rulemodel_output}:

% We now mention the differentiating details for each of the five constructed models.  We pair this with a discussion of the results in Table \ref{tab:rulemodel_output} and how they reflect the nature of each condition. 

\begin{table}[]
\caption{Performance of Rule-Based Models on 2,000 test instances, for different justification conditions on 10 training instances. Because the test dataset is balanced, the Recall (R) value is equivalent to Accuracy.  The last three columns are the really meaningful ones, as they highlight the absolute differences in Precision/Recall/F1 between the two classes (lower is better; values below 0.10 are highlighted). The Trivial model, which always assigns a positive label to each instance, is shown for reference. }
\begin{tabular}{llllllllll}
\toprule
 & \multicolumn{3}{l}{\textbf{Positive Class}} & \multicolumn{3}{l}{\textbf{Negative Class}} & \multicolumn{3}{l}{\textbf{Delta Between Classes}} \\
 \midrule
 & \textbf{P} & \textbf{R} & \textbf{F} & \textbf{P} & \textbf{R} & \textbf{F} & \textbf{P} & \textbf{R} & \textbf{F} \\
 \midrule
\textit{\textbf{Trivial (Always Pos)}} & \textit{0.5} & \textit{1.0} & \textit{0.667} & \textit{0.0} & \textit{0.0} & \textit{0.0} & 0.5 & 1.0 & 0.667 \\
\textbf{Bag of Words} & 0.641 & 0.886 & 0.744 & 0.968 & 0.03 & 0.058 & 0.327 & 0.856 & 0.686 \\
\textbf{Perturbation} & 0.768 & 0.076 & 0.138 & 0.891 & 0.041 & 0.078 & 0.123 & \textbf{0.035} & \textbf{0.060} \\
\textbf{Simplification} & 0.775 & 0.069 & 0.127 & 0.857 & 0.030 & 0.058 & \textbf{0.082} & \textbf{0.039} & \textbf{0.069} \\
\textbf{Concept Bag of Words} & 0.735 & 0.219 & 0.337 & 0.836 & 0.102 & 0.182 & 0.101 & 0.117 & 0.155 \\
\textbf{Concept Annotation} & 0.723 & 0.245 & 0.366 & 0.806 & 0.112 & 0.197 & \textbf{0.083} & 0.133 & 0.169 \\
\bottomrule
\end{tabular}
\label{tab:rulemodel_output}
\vspace{-5mm}
\end{table}

\textbf{Bag of words.} This was simple keyword matching on the terms identified in this condition. The positive terms output from this condition were mostly generic (`amazing', `delicious') whereas many negative terms were very specific (`over-hyped',`small quantities'). This is an artifact of both the domain (restaurant reviews) and the labels. The performance on the two classes reflects this: the positive class has pretty bad precision but great recall, as it severely over-generalizes, whereas the negative class has amazing precision but barely finds any examples, because  it is so specific. 

\textbf{Perturbation.} The perturbed parts of the instances were treated as local high quality training instances for both labels. All possible `Noun is Adjective' signals were extracted from those instances to populate the relevant dictionaries. If a verb was negated, or an adjective transformed into an antonym (e.g., changing `delicious' to `disgusting' in `There were delicious burgers', assigned a positive label), this meant that the topic (`burgers') is highly relevant, the original text (`delicious burgers') was a good training example for the given label, and the perturbed result (`disgusting burgers') was a good example for the opposite label.

\textbf{Simplification.} The simplified instances were treated as high quality training instances. All possible 'Noun is Adjective' signals were extracted from those instances to populate the relevant dictionaries. These signals did not overlap much in content, so the model could do little generalizing. Much like the \just{perturbation} condition, the recall for both classes is therefore extremely low, and the precision is respectable for only 10 training examples. \just{Perturbation} recall results are slightly better because each perturbed instance yields both a positive and a negative signal.

\textbf{Concept bag of words and Concept annotation} The concept taxonomy described in Section \ref{subsection:crowd_experiments} follows the `Noun is Adjective' format by definition, so it was encoded accordingly for both of these conditions. The outputs of each condition were then used to extend the possible dictionaries. For \textit{concept bag of words}, each annotation was added to both the `Noun' and `Adjective' dictionaries (whenever grammatically possible). For \textit{concept annotation}, the `Noun' and `Adjective' elements were elicited separately, and thus were added to their respective dictionaries. It is unclear that either of these conditions is more successful than the other, at this stage. The recall is markedly better than for \textit{simplification} and \textit{perturbation} owing to the well-structured concept taxonomy, that lends itself well to generalization. But this comes at a price, as the delta in performance between the classes is similarly worse.

\section{Discussion}

% \begin{itemize}
% \item used in broader domain, evidence? 
% \item Ziva helps saving domain experts time. 
%     \item speculation on elicitation -> not actually simplest
%     \item comparison to AutoAI
%     \VL{Why autoAI? I am not seeing an obvious connection}\SP{I was wondering Ziva can be integrated into AutoAI. Currently, in AutoAI, problem statement and domain-knowledge learning part is still not addressed. With Ziva, domain experts can provide a little domain knowledge and may be from then, they can automatically create an ML model}
    
%     \item how to collect nuanced domain knowledge - future work? domain knowledge that ziva can't express
%     \item workload on domain experts side - is it enough to review only 10 data inputs? How to set that N? 
%     \item combine different justification techniques
% \end{itemize}

% \textit{Use in broader domains}? Ziva helps saving domain experts time.} \SP{todo for marina, fyi I also mentioned restaurant domain below as a limitation of our paper. If you want to combine this paragraph with it}

\textbf{Capturing nuanced domain knowledge.} 
While Ziva provides basic components in a domain, data scientists pointed out there is some in that the current design of Ziva does not provide. For instance, domain experts provide insight about data, such as sparsity of a certain column. Data scientists find such information helpful, however, it can not be captured in the Ziva output. More investigation is required on how to extract those nuanced data. One possible direction is to leverage proposed documentation for data ~\cite{gebru2018datasheets} or for models ~\cite{arnold2019factsheets, mitchell2019model}. Another tactic is to take a set of guided questions similar to the ones proposed in ~\cite{saleh2020clinical} in discussions between domain experts and data scientists. The structure provided by these artifacts can facilitate transfer of domain knowledge and get teams to a common ground quickly.

\textbf{Re-evaluating the old normal: Bag of words.} 
Bag of words is one of the dominant ways in NLP domain to elicit signals.
It is seemingly most simple and straightforward task for domain experts. However, to our surprise, our work suggests otherwise. Participants in our user study indicated that \just{bag of words} is in fact more mentally demanding, harder to accomplish and more stressful for participants than other \name{justification} techniques. Furthermore, in our exercise of building a rule-based model with different \name{justification} methods, the other methods outperformed the \just{bag of words}. This informed us that both domain experts and data scientists can benefit from our \name{justification} techniques during collaboration. We believe our \name{justification} method could be used throughout the ML development workflow and provide an outlet for stakeholders to efficiently communicate their model building. 

\textbf{Limitations.} 
%So far, we have only conducted a case study on a Yelp sentiment classifier. 
Various use cases of Ziva output validated the efficacy of our interfaces drawn from our preliminary study and literature review, demonstrating that domain experts' elicited knowledge can facilitate model building. However, this paper only considers the concrete setting of a sentiment classifier for Yelp restaurant reviews. 
\rev{Future work should examine the generalizability of the approach for other tasks (e.g., document classification, clustering, machine translation, and question answering) and other domains (e.g., education, health science).}
Nevertheless, the overall approach described in this paper is domain-agnostic, and in fact much more relevant to real-life scenarios with complex tasks, specialized domains, and significant constraints on the resources to generate large amounts of labeled data. We 
%Regardless of the fact that Ziva is tested on the Yelp dataset, we 
therefore believe we have identified a number of interesting design requirements of domain-knowledge sharing in ML development workflow that are not currently addressed, and are applicable across tasks and domains.

\section{Conclusion}

In this paper, we presented a system and a case study on how data scientists can get help from domain experts in ML development lifecycle. Along the way, we were able to identify the current practice of how data scientists go about domain knowledge-learning. Inspired by the workaround to extract domain knowledge, we designed an interface that facilitates the domain knowledge-sharing. We presented the interface output to ML practitioners to reflect their experience building an ML model in a specialized domain. They shared that scalability of a piece of domain knowledge and low cognitive load of domain experts are important factors in such a domain knowledge-bootstrapping tool. We continued the work by investigating cognitive load of different methods in our interface. We found that the traditional elicitation method ``bag of words'' is least preferred by domain experts in terms of mental load and stress level, and provides the least knowledge scalability compared to other elicitation methods.    

%%
%% The acknowledgments section is defined using the "acks" environment
%% (and NOT an unnumbered section). This ensures the proper
%% identification of the section in the article metadata, and the
%% consistent spelling of the heading.
\begin{acks}
We thank Dakuo Wang and David Karger for their feedback. Soya Park is partly supported by the Kwanjeong fellowship.
\end{acks}

%%
%% The next two lines define the bibliography style to be used, and
%% the bibliography file.
\bibliographystyle{ACM-Reference-Format}
\bibliography{sample-base}

%%% -*-BibTeX-*-
%%% Do NOT edit. File created by BibTeX with style
%%% ACM-Reference-Format-Journals [18-Jan-2012].

\begin{thebibliography}{91}

%%% ====================================================================
%%% NOTE TO THE USER: you can override these defaults by providing
%%% customized versions of any of these macros before the \bibliography
%%% command.  Each of them MUST provide its own final punctuation,
%%% except for \shownote{}, \showDOI{}, and \showURL{}.  The latter two
%%% do not use final punctuation, in order to avoid confusing it with
%%% the Web address.
%%%
%%% To suppress output of a particular field, define its macro to expand
%%% to an empty string, or better, \unskip, like this:
%%%
%%% \newcommand{\showDOI}[1]{\unskip}   % LaTeX syntax
%%%
%%% \def \showDOI #1{\unskip}           % plain TeX syntax
%%%
%%% ====================================================================

\ifx \showCODEN    \undefined \def \showCODEN     #1{\unskip}     \fi
\ifx \showDOI      \undefined \def \showDOI       #1{#1}\fi
\ifx \showISBNx    \undefined \def \showISBNx     #1{\unskip}     \fi
\ifx \showISBNxiii \undefined \def \showISBNxiii  #1{\unskip}     \fi
\ifx \showISSN     \undefined \def \showISSN      #1{\unskip}     \fi
\ifx \showLCCN     \undefined \def \showLCCN      #1{\unskip}     \fi
\ifx \shownote     \undefined \def \shownote      #1{#1}          \fi
\ifx \showarticletitle \undefined \def \showarticletitle #1{#1}   \fi
\ifx \showURL      \undefined \def \showURL       {\relax}        \fi
% The following commands are used for tagged output and should be
% invisible to TeX
\providecommand\bibfield[2]{#2}
\providecommand\bibinfo[2]{#2}
\providecommand\natexlab[1]{#1}
\providecommand\showeprint[2][]{arXiv:#2}

\bibitem[\protect\citeauthoryear{??}{app}{2021}]%
        {appen}
 \bibinfo{year}{2021}\natexlab{}.
\newblock \bibinfo{title}{Appen}.
\newblock \bibinfo{howpublished}{\url{https://appen.com}}.
\newblock


\bibitem[\protect\citeauthoryear{??}{lmd}{2021}]%
        {lmdd}
 \bibinfo{year}{2021}\natexlab{}.
\newblock \bibinfo{title}{Lean-Mean-Drag-and-Drop}.
\newblock
  \bibinfo{howpublished}{\url{https://supraniti.github.io/Lean-Mean-Drag-and-Drop/}}.
\newblock


\bibitem[\protect\citeauthoryear{??}{yel}{2021}]%
        {yelp}
 \bibinfo{year}{2021}\natexlab{}.
\newblock \bibinfo{title}{Yelp Open Dataset}.
\newblock \bibinfo{howpublished}{\url{https://www.yelp.com/dataset}}.
\newblock


\bibitem[\protect\citeauthoryear{Ackerman}{Ackerman}{1998}]%
        {ackerman1998augmenting}
\bibfield{author}{\bibinfo{person}{Mark~S Ackerman}.}
  \bibinfo{year}{1998}\natexlab{}.
\newblock \showarticletitle{Augmenting organizational memory: a field study of
  answer garden}.
\newblock \bibinfo{journal}{\emph{ACM Transactions on Information Systems
  (TOIS)}} \bibinfo{volume}{16}, \bibinfo{number}{3} (\bibinfo{year}{1998}),
  \bibinfo{pages}{203--224}.
\newblock


\bibitem[\protect\citeauthoryear{Ackerman, Dachtera, Pipek, and Wulf}{Ackerman
  et~al\mbox{.}}{2013}]%
        {ackerman2013sharing}
\bibfield{author}{\bibinfo{person}{Mark~S Ackerman}, \bibinfo{person}{Juri
  Dachtera}, \bibinfo{person}{Volkmar Pipek}, {and} \bibinfo{person}{Volker
  Wulf}.} \bibinfo{year}{2013}\natexlab{}.
\newblock \showarticletitle{Sharing knowledge and expertise: The CSCW view of
  knowledge management}.
\newblock \bibinfo{journal}{\emph{Computer Supported Cooperative Work (CSCW)}}
  \bibinfo{volume}{22}, \bibinfo{number}{4-6} (\bibinfo{year}{2013}),
  \bibinfo{pages}{531--573}.
\newblock


\bibitem[\protect\citeauthoryear{Akbik and Li}{Akbik and Li}{2016}]%
        {Akbik2016KSRLIL}
\bibfield{author}{\bibinfo{person}{A. Akbik} {and} \bibinfo{person}{Yunyao
  Li}.} \bibinfo{year}{2016}\natexlab{}.
\newblock \showarticletitle{K-SRL: Instance-based Learning for Semantic Role
  Labeling}. In \bibinfo{booktitle}{\emph{COLING}}.
\newblock


\bibitem[\protect\citeauthoryear{Alzantot, Sharma, Elgohary, Ho, Srivastava,
  and Chang}{Alzantot et~al\mbox{.}}{2018}]%
        {alzantot-etal-2018-generating}
\bibfield{author}{\bibinfo{person}{Moustafa Alzantot}, \bibinfo{person}{Yash
  Sharma}, \bibinfo{person}{Ahmed Elgohary}, \bibinfo{person}{Bo-Jhang Ho},
  \bibinfo{person}{Mani Srivastava}, {and} \bibinfo{person}{Kai-Wei Chang}.}
  \bibinfo{year}{2018}\natexlab{}.
\newblock \showarticletitle{Generating Natural Language Adversarial Examples}.
  In \bibinfo{booktitle}{\emph{Proceedings of the 2018 Conference on Empirical
  Methods in Natural Language Processing}}. \bibinfo{publisher}{Association for
  Computational Linguistics}, \bibinfo{address}{Brussels, Belgium},
  \bibinfo{pages}{2890--2896}.
\newblock
\urldef\tempurl%
\url{https://doi.org/10.18653/v1/D18-1316}
\showDOI{\tempurl}


\bibitem[\protect\citeauthoryear{Amershi, Begel, Bird, DeLine, Gall, Kamar,
  Nagappan, Nushi, and Zimmermann}{Amershi et~al\mbox{.}}{2019}]%
        {amershi2019software}
\bibfield{author}{\bibinfo{person}{Saleema Amershi}, \bibinfo{person}{Andrew
  Begel}, \bibinfo{person}{Christian Bird}, \bibinfo{person}{Robert DeLine},
  \bibinfo{person}{Harald Gall}, \bibinfo{person}{Ece Kamar},
  \bibinfo{person}{Nachiappan Nagappan}, \bibinfo{person}{Besmira Nushi}, {and}
  \bibinfo{person}{Thomas Zimmermann}.} \bibinfo{year}{2019}\natexlab{}.
\newblock \showarticletitle{Software engineering for machine learning: A case
  study}. In \bibinfo{booktitle}{\emph{2019 IEEE/ACM 41st International
  Conference on Software Engineering: Software Engineering in Practice
  (ICSE-SEIP)}}. IEEE, \bibinfo{pages}{291--300}.
\newblock


\bibitem[\protect\citeauthoryear{Amershi, Cakmak, Knox, and Kulesza}{Amershi
  et~al\mbox{.}}{2014}]%
        {amershi2014power}
\bibfield{author}{\bibinfo{person}{Saleema Amershi}, \bibinfo{person}{Maya
  Cakmak}, \bibinfo{person}{William~Bradley Knox}, {and} \bibinfo{person}{Todd
  Kulesza}.} \bibinfo{year}{2014}\natexlab{}.
\newblock \showarticletitle{Power to the people: The role of humans in
  interactive machine learning}.
\newblock \bibinfo{journal}{\emph{Ai Magazine}} \bibinfo{volume}{35},
  \bibinfo{number}{4} (\bibinfo{year}{2014}), \bibinfo{pages}{105--120}.
\newblock


\bibitem[\protect\citeauthoryear{Amershi, Chickering, Drucker, Lee, Simard, and
  Suh}{Amershi et~al\mbox{.}}{2015}]%
        {amershi2015modeltracker}
\bibfield{author}{\bibinfo{person}{Saleema Amershi}, \bibinfo{person}{Max
  Chickering}, \bibinfo{person}{Steven~M Drucker}, \bibinfo{person}{Bongshin
  Lee}, \bibinfo{person}{Patrice Simard}, {and} \bibinfo{person}{Jina Suh}.}
  \bibinfo{year}{2015}\natexlab{}.
\newblock \showarticletitle{Modeltracker: Redesigning performance analysis
  tools for machine learning}. In \bibinfo{booktitle}{\emph{Proceedings of the
  33rd Annual ACM Conference on Human Factors in Computing Systems}}.
  \bibinfo{pages}{337--346}.
\newblock


\bibitem[\protect\citeauthoryear{Andrzejewski, Zhu, and Craven}{Andrzejewski
  et~al\mbox{.}}{2009}]%
        {andrzejewski2009incorporating}
\bibfield{author}{\bibinfo{person}{David Andrzejewski},
  \bibinfo{person}{Xiaojin Zhu}, {and} \bibinfo{person}{Mark Craven}.}
  \bibinfo{year}{2009}\natexlab{}.
\newblock \showarticletitle{Incorporating domain knowledge into topic modeling
  via Dirichlet forest priors}. In \bibinfo{booktitle}{\emph{Proceedings of the
  26th annual international conference on machine learning}}.
  \bibinfo{pages}{25--32}.
\newblock


\bibitem[\protect\citeauthoryear{Arnold, Bellamy, Hind, Houde, Mehta,
  Mojsilovi{\'c}, Nair, Ramamurthy, Olteanu, Piorkowski, et~al\mbox{.}}{Arnold
  et~al\mbox{.}}{2019}]%
        {arnold2019factsheets}
\bibfield{author}{\bibinfo{person}{Matthew Arnold}, \bibinfo{person}{Rachel~KE
  Bellamy}, \bibinfo{person}{Michael Hind}, \bibinfo{person}{Stephanie Houde},
  \bibinfo{person}{Sameep Mehta}, \bibinfo{person}{A Mojsilovi{\'c}},
  \bibinfo{person}{Ravi Nair}, \bibinfo{person}{K~Natesan Ramamurthy},
  \bibinfo{person}{Alexandra Olteanu}, \bibinfo{person}{David Piorkowski},
  {et~al\mbox{.}}} \bibinfo{year}{2019}\natexlab{}.
\newblock \showarticletitle{FactSheets: Increasing trust in AI services through
  supplier's declarations of conformity}.
\newblock \bibinfo{journal}{\emph{IBM Journal of Research and Development}}
  \bibinfo{volume}{63}, \bibinfo{number}{4/5} (\bibinfo{year}{2019}),
  \bibinfo{pages}{6--1}.
\newblock


\bibitem[\protect\citeauthoryear{Bao, Wu, Chang, and Barzilay}{Bao
  et~al\mbox{.}}{2019}]%
        {bao2019few}
\bibfield{author}{\bibinfo{person}{Yujia Bao}, \bibinfo{person}{Menghua Wu},
  \bibinfo{person}{Shiyu Chang}, {and} \bibinfo{person}{Regina Barzilay}.}
  \bibinfo{year}{2019}\natexlab{}.
\newblock \showarticletitle{Few-shot text classification with distributional
  signatures}.
\newblock \bibinfo{journal}{\emph{arXiv preprint arXiv:1908.06039}}
  (\bibinfo{year}{2019}).
\newblock


\bibitem[\protect\citeauthoryear{Brooks, Amershi, Lee, Drucker, Kapoor, and
  Simard}{Brooks et~al\mbox{.}}{2015}]%
        {brooks2015featureinsight}
\bibfield{author}{\bibinfo{person}{Michael Brooks}, \bibinfo{person}{Saleema
  Amershi}, \bibinfo{person}{Bongshin Lee}, \bibinfo{person}{Steven~M Drucker},
  \bibinfo{person}{Ashish Kapoor}, {and} \bibinfo{person}{Patrice Simard}.}
  \bibinfo{year}{2015}\natexlab{}.
\newblock \showarticletitle{FeatureInsight: Visual support for error-driven
  feature ideation in text classification}. In \bibinfo{booktitle}{\emph{2015
  IEEE Conference on Visual Analytics Science and Technology (VAST)}}. IEEE,
  \bibinfo{pages}{105--112}.
\newblock


\bibitem[\protect\citeauthoryear{Cai and Guo}{Cai and Guo}{2019}]%
        {cai2019software}
\bibfield{author}{\bibinfo{person}{Carrie~J Cai} {and}
  \bibinfo{person}{Philip~J Guo}.} \bibinfo{year}{2019}\natexlab{}.
\newblock \showarticletitle{Software Developers Learning Machine Learning:
  Motivations, Hurdles, and Desires}. In \bibinfo{booktitle}{\emph{2019 IEEE
  Symposium on Visual Languages and Human-Centric Computing (VL/HCC)}}. IEEE,
  \bibinfo{pages}{25--34}.
\newblock


\bibitem[\protect\citeauthoryear{Cai, Reif, Hegde, Hipp, Kim, Smilkov,
  Wattenberg, Viégas, Corrado, Stumpe, and Terry}{Cai et~al\mbox{.}}{2019a}]%
        {47823}
\bibfield{author}{\bibinfo{person}{Carrie~Jun Cai}, \bibinfo{person}{Emily
  Reif}, \bibinfo{person}{Narayan~G Hegde}, \bibinfo{person}{Jason Hipp},
  \bibinfo{person}{Been Kim}, \bibinfo{person}{Daniel Smilkov},
  \bibinfo{person}{Martin Wattenberg}, \bibinfo{person}{Fernanda Viégas},
  \bibinfo{person}{Greg Corrado}, \bibinfo{person}{Martin Stumpe}, {and}
  \bibinfo{person}{Michael Terry}.} \bibinfo{year}{2019}\natexlab{a}.
\newblock \showarticletitle{Human-Centered Tools for Coping with Imperfect
  Algorithms during Medical Decision-Making}.
\newblock
\urldef\tempurl%
\url{https://arxiv.org/abs/1902.02960}
\showURL{%
\tempurl}


\bibitem[\protect\citeauthoryear{Cai, Winter, Steiner, Wilcox, and Terry}{Cai
  et~al\mbox{.}}{2019b}]%
        {48431}
\bibfield{author}{\bibinfo{person}{Carrie~Jun Cai}, \bibinfo{person}{Samantha
  Winter}, \bibinfo{person}{David Steiner}, \bibinfo{person}{Lauren Wilcox},
  {and} \bibinfo{person}{Michael Terry}.} \bibinfo{year}{2019}\natexlab{b}.
\newblock \showarticletitle{"Hello AI": Uncovering the Onboarding Needs of
  Medical Practitioners for Human-AI Collaborative Decision-Making}.
\newblock


\bibitem[\protect\citeauthoryear{Cakmak and Thomaz}{Cakmak and Thomaz}{2012}]%
        {cakmak2012designing}
\bibfield{author}{\bibinfo{person}{Maya Cakmak} {and} \bibinfo{person}{Andrea~L
  Thomaz}.} \bibinfo{year}{2012}\natexlab{}.
\newblock \showarticletitle{Designing robot learners that ask good questions}.
  In \bibinfo{booktitle}{\emph{2012 7th ACM/IEEE International Conference on
  Human-Robot Interaction (HRI)}}. IEEE, \bibinfo{pages}{17--24}.
\newblock


\bibitem[\protect\citeauthoryear{Chang, Ratinov, and Roth}{Chang
  et~al\mbox{.}}{2007}]%
        {chang2007guiding}
\bibfield{author}{\bibinfo{person}{Ming-Wei Chang}, \bibinfo{person}{Lev
  Ratinov}, {and} \bibinfo{person}{Dan Roth}.} \bibinfo{year}{2007}\natexlab{}.
\newblock \showarticletitle{Guiding semi-supervision with constraint-driven
  learning}. In \bibinfo{booktitle}{\emph{Proceedings of the 45th annual
  meeting of the association of computational linguistics}}.
  \bibinfo{pages}{280--287}.
\newblock


\bibitem[\protect\citeauthoryear{Chilton, Little, Edge, Weld, and
  Landay}{Chilton et~al\mbox{.}}{2013}]%
        {chilton2013cascade}
\bibfield{author}{\bibinfo{person}{Lydia~B Chilton}, \bibinfo{person}{Greg
  Little}, \bibinfo{person}{Darren Edge}, \bibinfo{person}{Daniel~S Weld},
  {and} \bibinfo{person}{James~A Landay}.} \bibinfo{year}{2013}\natexlab{}.
\newblock \showarticletitle{Cascade: Crowdsourcing taxonomy creation}. In
  \bibinfo{booktitle}{\emph{Proceedings of the SIGCHI Conference on Human
  Factors in Computing Systems}}. \bibinfo{pages}{1999--2008}.
\newblock


\bibitem[\protect\citeauthoryear{Choo, Lee, Reddy, and Park}{Choo
  et~al\mbox{.}}{2013}]%
        {choo2013utopian}
\bibfield{author}{\bibinfo{person}{Jaegul Choo}, \bibinfo{person}{Changhyun
  Lee}, \bibinfo{person}{Chandan~K Reddy}, {and} \bibinfo{person}{Haesun
  Park}.} \bibinfo{year}{2013}\natexlab{}.
\newblock \showarticletitle{Utopian: User-driven topic modeling based on
  interactive nonnegative matrix factorization}.
\newblock \bibinfo{journal}{\emph{IEEE transactions on visualization and
  computer graphics}} \bibinfo{volume}{19}, \bibinfo{number}{12}
  (\bibinfo{year}{2013}), \bibinfo{pages}{1992--2001}.
\newblock


\bibitem[\protect\citeauthoryear{Crowston, Saltz, Rezgui, Hegde, and
  You}{Crowston et~al\mbox{.}}{2019}]%
        {crowston2019socio}
\bibfield{author}{\bibinfo{person}{Kevin Crowston}, \bibinfo{person}{Jeff~S
  Saltz}, \bibinfo{person}{Amira Rezgui}, \bibinfo{person}{Yatish Hegde}, {and}
  \bibinfo{person}{Sangseok You}.} \bibinfo{year}{2019}\natexlab{}.
\newblock \showarticletitle{Socio-technical Affordances for Stigmergic
  Coordination Implemented in MIDST, a Tool for Data-Science Teams}.
\newblock \bibinfo{journal}{\emph{Proceedings of the ACM on Human-Computer
  Interaction}} \bibinfo{volume}{3}, \bibinfo{number}{CSCW}
  (\bibinfo{year}{2019}), \bibinfo{pages}{1--25}.
\newblock


\bibitem[\protect\citeauthoryear{Culkin and Das}{Culkin and Das}{2017}]%
        {culkin2017machine}
\bibfield{author}{\bibinfo{person}{Robert Culkin} {and}
  \bibinfo{person}{Sanjiv~R Das}.} \bibinfo{year}{2017}\natexlab{}.
\newblock \showarticletitle{Machine learning in finance: The case of deep
  learning for option pricing}.
\newblock \bibinfo{journal}{\emph{Journal of Investment Management}}
  \bibinfo{volume}{15}, \bibinfo{number}{4} (\bibinfo{year}{2017}),
  \bibinfo{pages}{92--100}.
\newblock


\bibitem[\protect\citeauthoryear{Dehghani, Zamani, Severyn, Kamps, and
  Croft}{Dehghani et~al\mbox{.}}{2017}]%
        {dehghani2017neural}
\bibfield{author}{\bibinfo{person}{Mostafa Dehghani}, \bibinfo{person}{Hamed
  Zamani}, \bibinfo{person}{Aliaksei Severyn}, \bibinfo{person}{Jaap Kamps},
  {and} \bibinfo{person}{W~Bruce Croft}.} \bibinfo{year}{2017}\natexlab{}.
\newblock \showarticletitle{Neural ranking models with weak supervision}. In
  \bibinfo{booktitle}{\emph{Proceedings of the 40th International ACM SIGIR
  Conference on Research and Development in Information Retrieval}}.
  \bibinfo{pages}{65--74}.
\newblock


\bibitem[\protect\citeauthoryear{Druck, Settles, and McCallum}{Druck
  et~al\mbox{.}}{2009}]%
        {druck2009active}
\bibfield{author}{\bibinfo{person}{Gregory Druck}, \bibinfo{person}{Burr
  Settles}, {and} \bibinfo{person}{Andrew McCallum}.}
  \bibinfo{year}{2009}\natexlab{}.
\newblock \showarticletitle{Active learning by labeling features}. In
  \bibinfo{booktitle}{\emph{Proceedings of the 2009 conference on Empirical
  methods in natural language processing}}. \bibinfo{pages}{81--90}.
\newblock


\bibitem[\protect\citeauthoryear{Fails and Olsen~Jr}{Fails and
  Olsen~Jr}{2003}]%
        {fails2003interactive}
\bibfield{author}{\bibinfo{person}{Jerry~Alan Fails} {and}
  \bibinfo{person}{Dan~R Olsen~Jr}.} \bibinfo{year}{2003}\natexlab{}.
\newblock \showarticletitle{Interactive machine learning}. In
  \bibinfo{booktitle}{\emph{Proceedings of the 8th international conference on
  Intelligent user interfaces}}. \bibinfo{pages}{39--45}.
\newblock


\bibitem[\protect\citeauthoryear{Fogarty, Tan, Kapoor, and Winder}{Fogarty
  et~al\mbox{.}}{2008}]%
        {fogarty2008cueflik}
\bibfield{author}{\bibinfo{person}{James Fogarty}, \bibinfo{person}{Desney
  Tan}, \bibinfo{person}{Ashish Kapoor}, {and} \bibinfo{person}{Simon Winder}.}
  \bibinfo{year}{2008}\natexlab{}.
\newblock \showarticletitle{CueFlik: interactive concept learning in image
  search}. In \bibinfo{booktitle}{\emph{Proceedings of the sigchi conference on
  human factors in computing systems}}. \bibinfo{pages}{29--38}.
\newblock


\bibitem[\protect\citeauthoryear{Gebru, Morgenstern, Vecchione, Vaughan,
  Wallach, Daum{\'e}~III, and Crawford}{Gebru et~al\mbox{.}}{2018}]%
        {gebru2018datasheets}
\bibfield{author}{\bibinfo{person}{Timnit Gebru}, \bibinfo{person}{Jamie
  Morgenstern}, \bibinfo{person}{Briana Vecchione},
  \bibinfo{person}{Jennifer~Wortman Vaughan}, \bibinfo{person}{Hanna Wallach},
  \bibinfo{person}{Hal Daum{\'e}~III}, {and} \bibinfo{person}{Kate Crawford}.}
  \bibinfo{year}{2018}\natexlab{}.
\newblock \showarticletitle{Datasheets for datasets}.
\newblock \bibinfo{journal}{\emph{arXiv preprint arXiv:1803.09010}}
  (\bibinfo{year}{2018}).
\newblock


\bibitem[\protect\citeauthoryear{Ghai, Liao, Zhang, Bellamy, and Mueller}{Ghai
  et~al\mbox{.}}{2020}]%
        {ghai2020explainable}
\bibfield{author}{\bibinfo{person}{Bhavya Ghai}, \bibinfo{person}{Q~Vera Liao},
  \bibinfo{person}{Yunfeng Zhang}, \bibinfo{person}{Rachel Bellamy}, {and}
  \bibinfo{person}{Klaus Mueller}.} \bibinfo{year}{2020}\natexlab{}.
\newblock \showarticletitle{Explainable Active Learning (XAL): An Empirical
  Study of How Local Explanations Impact Annotator Experience}.
\newblock \bibinfo{journal}{\emph{arXiv preprint arXiv:2001.09219}}
  (\bibinfo{year}{2020}).
\newblock


\bibitem[\protect\citeauthoryear{Ghorbani, Wexler, Zou, and Kim}{Ghorbani
  et~al\mbox{.}}{2019}]%
        {ghorbani2019towards}
\bibfield{author}{\bibinfo{person}{Amirata Ghorbani}, \bibinfo{person}{James
  Wexler}, \bibinfo{person}{James~Y Zou}, {and} \bibinfo{person}{Been Kim}.}
  \bibinfo{year}{2019}\natexlab{}.
\newblock \showarticletitle{Towards automatic concept-based explanations}. In
  \bibinfo{booktitle}{\emph{Advances in Neural Information Processing
  Systems}}. \bibinfo{pages}{9273--9282}.
\newblock


\bibitem[\protect\citeauthoryear{Goldberg, Safran, and Shapiro}{Goldberg
  et~al\mbox{.}}{1992}]%
        {goldberg1992active}
\bibfield{author}{\bibinfo{person}{Yaron Goldberg}, \bibinfo{person}{Marilyn
  Safran}, {and} \bibinfo{person}{Ehud Shapiro}.}
  \bibinfo{year}{1992}\natexlab{}.
\newblock \showarticletitle{Active mail—a framework for implementing
  groupware}. In \bibinfo{booktitle}{\emph{Proceedings of the 1992 ACM
  conference on Computer-supported cooperative work}}. \bibinfo{pages}{75--83}.
\newblock


\bibitem[\protect\citeauthoryear{Grudin}{Grudin}{1988}]%
        {grudin1988cscw}
\bibfield{author}{\bibinfo{person}{Jonathan Grudin}.}
  \bibinfo{year}{1988}\natexlab{}.
\newblock \showarticletitle{Why CSCW applications fail: problems in the design
  and evaluationof organizational interfaces}. In
  \bibinfo{booktitle}{\emph{Proceedings of the 1988 ACM conference on
  Computer-supported cooperative work}}. \bibinfo{pages}{85--93}.
\newblock


\bibitem[\protect\citeauthoryear{Hart and Staveland}{Hart and
  Staveland}{1988}]%
        {hart1988development}
\bibfield{author}{\bibinfo{person}{Sandra~G Hart} {and}
  \bibinfo{person}{Lowell~E Staveland}.} \bibinfo{year}{1988}\natexlab{}.
\newblock \showarticletitle{Development of NASA-TLX (Task Load Index): Results
  of empirical and theoretical research}.
\newblock In \bibinfo{booktitle}{\emph{Advances in psychology}}.
  Vol.~\bibinfo{volume}{52}. \bibinfo{publisher}{Elsevier},
  \bibinfo{pages}{139--183}.
\newblock


\bibitem[\protect\citeauthoryear{Head, Hohman, Barik, Drucker, and DeLine}{Head
  et~al\mbox{.}}{2019}]%
        {head2019managing}
\bibfield{author}{\bibinfo{person}{Andrew Head}, \bibinfo{person}{Fred Hohman},
  \bibinfo{person}{Titus Barik}, \bibinfo{person}{Steven~M Drucker}, {and}
  \bibinfo{person}{Robert DeLine}.} \bibinfo{year}{2019}\natexlab{}.
\newblock \showarticletitle{Managing messes in computational notebooks}. In
  \bibinfo{booktitle}{\emph{Proceedings of the 2019 CHI Conference on Human
  Factors in Computing Systems}}. \bibinfo{pages}{1--12}.
\newblock


\bibitem[\protect\citeauthoryear{Hoffmann, de~Carvalho, Abele, Schweitzer,
  Tolmie, and Wulf}{Hoffmann et~al\mbox{.}}{2019}]%
        {hoffmann2019cyber}
\bibfield{author}{\bibinfo{person}{Sven Hoffmann}, \bibinfo{person}{Aparecido
  Fabiano~Pinatti de Carvalho}, \bibinfo{person}{Darwin Abele},
  \bibinfo{person}{Marcus Schweitzer}, \bibinfo{person}{Peter Tolmie}, {and}
  \bibinfo{person}{Volker Wulf}.} \bibinfo{year}{2019}\natexlab{}.
\newblock \showarticletitle{Cyber-Physical Systems for Knowledge and Expertise
  Sharing in Manufacturing Contexts: Towards a Model Enabling Design}.
\newblock \bibinfo{journal}{\emph{Computer Supported Cooperative Work (CSCW)}}
  \bibinfo{volume}{28}, \bibinfo{number}{3-4} (\bibinfo{year}{2019}),
  \bibinfo{pages}{469--509}.
\newblock


\bibitem[\protect\citeauthoryear{Hohman, Head, Caruana, DeLine, and
  Drucker}{Hohman et~al\mbox{.}}{2019}]%
        {hohman2019gamut}
\bibfield{author}{\bibinfo{person}{Fred Hohman}, \bibinfo{person}{Andrew Head},
  \bibinfo{person}{Rich Caruana}, \bibinfo{person}{Robert DeLine}, {and}
  \bibinfo{person}{Steven~M Drucker}.} \bibinfo{year}{2019}\natexlab{}.
\newblock \showarticletitle{Gamut: A design probe to understand how data
  scientists understand machine learning models}. In
  \bibinfo{booktitle}{\emph{Proceedings of the 2019 CHI conference on human
  factors in computing systems}}. \bibinfo{pages}{1--13}.
\newblock


\bibitem[\protect\citeauthoryear{Hohman, Kahng, Pienta, and Chau}{Hohman
  et~al\mbox{.}}{2018}]%
        {hohman2018visual}
\bibfield{author}{\bibinfo{person}{Fred Hohman}, \bibinfo{person}{Minsuk
  Kahng}, \bibinfo{person}{Robert Pienta}, {and} \bibinfo{person}{Duen~Horng
  Chau}.} \bibinfo{year}{2018}\natexlab{}.
\newblock \showarticletitle{Visual analytics in deep learning: An interrogative
  survey for the next frontiers}.
\newblock \bibinfo{journal}{\emph{IEEE transactions on visualization and
  computer graphics}} \bibinfo{volume}{25}, \bibinfo{number}{8}
  (\bibinfo{year}{2018}), \bibinfo{pages}{2674--2693}.
\newblock


\bibitem[\protect\citeauthoryear{Holzinger}{Holzinger}{2016}]%
        {holzinger2016interactive}
\bibfield{author}{\bibinfo{person}{Andreas Holzinger}.}
  \bibinfo{year}{2016}\natexlab{}.
\newblock \showarticletitle{Interactive machine learning for health
  informatics: when do we need the human-in-the-loop?}
\newblock \bibinfo{journal}{\emph{Brain Informatics}} \bibinfo{volume}{3},
  \bibinfo{number}{2} (\bibinfo{year}{2016}), \bibinfo{pages}{119--131}.
\newblock


\bibitem[\protect\citeauthoryear{Hoque and Carenini}{Hoque and
  Carenini}{2015}]%
        {hoque2015convisit}
\bibfield{author}{\bibinfo{person}{Enamul Hoque} {and}
  \bibinfo{person}{Giuseppe Carenini}.} \bibinfo{year}{2015}\natexlab{}.
\newblock \showarticletitle{Convisit: Interactive topic modeling for exploring
  asynchronous online conversations}. In \bibinfo{booktitle}{\emph{Proceedings
  of the 20th International Conference on Intelligent User Interfaces}}.
  \bibinfo{pages}{169--180}.
\newblock


\bibitem[\protect\citeauthoryear{Hu, Boyd-Graber, Satinoff, and Smith}{Hu
  et~al\mbox{.}}{2014}]%
        {hu2014interactive}
\bibfield{author}{\bibinfo{person}{Yuening Hu}, \bibinfo{person}{Jordan
  Boyd-Graber}, \bibinfo{person}{Brianna Satinoff}, {and}
  \bibinfo{person}{Alison Smith}.} \bibinfo{year}{2014}\natexlab{}.
\newblock \showarticletitle{Interactive topic modeling}.
\newblock \bibinfo{journal}{\emph{Machine learning}} \bibinfo{volume}{95},
  \bibinfo{number}{3} (\bibinfo{year}{2014}), \bibinfo{pages}{423--469}.
\newblock


\bibitem[\protect\citeauthoryear{Jiang, Liu, and Chen}{Jiang
  et~al\mbox{.}}{2019}]%
        {jiang2019recent}
\bibfield{author}{\bibinfo{person}{Liu Jiang}, \bibinfo{person}{Shixia Liu},
  {and} \bibinfo{person}{Changjian Chen}.} \bibinfo{year}{2019}\natexlab{}.
\newblock \showarticletitle{Recent research advances on interactive machine
  learning}.
\newblock \bibinfo{journal}{\emph{Journal of Visualization}}
  \bibinfo{volume}{22}, \bibinfo{number}{2} (\bibinfo{year}{2019}),
  \bibinfo{pages}{401--417}.
\newblock


\bibitem[\protect\citeauthoryear{Jing}{Jing}{2000}]%
        {Jing_2000}
\bibfield{author}{\bibinfo{person}{Hongyan Jing}.}
  \bibinfo{year}{2000}\natexlab{}.
\newblock \showarticletitle{Sentence reduction for automatic text
  summarization}. In \bibinfo{booktitle}{\emph{Proceedings of the sixth
  conference on Applied natural language processing}}
  \emph{(\bibinfo{series}{ANLC ’00})}. \bibinfo{publisher}{Association for
  Computational Linguistics}, \bibinfo{pages}{310–315}.
\newblock
\urldef\tempurl%
\url{https://doi.org/10.3115/974147.974190}
\showDOI{\tempurl}


\bibitem[\protect\citeauthoryear{Kapoor, Lee, Tan, and Horvitz}{Kapoor
  et~al\mbox{.}}{2010}]%
        {kapoor2010interactive}
\bibfield{author}{\bibinfo{person}{Ashish Kapoor}, \bibinfo{person}{Bongshin
  Lee}, \bibinfo{person}{Desney Tan}, {and} \bibinfo{person}{Eric Horvitz}.}
  \bibinfo{year}{2010}\natexlab{}.
\newblock \showarticletitle{Interactive optimization for steering machine
  classification}. In \bibinfo{booktitle}{\emph{Proceedings of the SIGCHI
  Conference on Human Factors in Computing Systems}}.
  \bibinfo{pages}{1343--1352}.
\newblock


\bibitem[\protect\citeauthoryear{Kim, Wattenberg, Gilmer, Cai, Wexler, Viegas,
  et~al\mbox{.}}{Kim et~al\mbox{.}}{2018}]%
        {kim2018interpretability}
\bibfield{author}{\bibinfo{person}{Been Kim}, \bibinfo{person}{Martin
  Wattenberg}, \bibinfo{person}{Justin Gilmer}, \bibinfo{person}{Carrie Cai},
  \bibinfo{person}{James Wexler}, \bibinfo{person}{Fernanda Viegas},
  {et~al\mbox{.}}} \bibinfo{year}{2018}\natexlab{}.
\newblock \showarticletitle{Interpretability beyond feature attribution:
  Quantitative testing with concept activation vectors (tcav)}. In
  \bibinfo{booktitle}{\emph{International conference on machine learning}}.
  PMLR, \bibinfo{pages}{2668--2677}.
\newblock


\bibitem[\protect\citeauthoryear{Kim, Zimmermann, DeLine, and Begel}{Kim
  et~al\mbox{.}}{2016}]%
        {kim2016emerging}
\bibfield{author}{\bibinfo{person}{Miryung Kim}, \bibinfo{person}{Thomas
  Zimmermann}, \bibinfo{person}{Robert DeLine}, {and} \bibinfo{person}{Andrew
  Begel}.} \bibinfo{year}{2016}\natexlab{}.
\newblock \showarticletitle{The emerging role of data scientists on software
  development teams}. In \bibinfo{booktitle}{\emph{Proceedings of the 38th
  International Conference on Software Engineering}}. ACM,
  \bibinfo{pages}{96--107}.
\newblock


\bibitem[\protect\citeauthoryear{Krause, Perer, and Bertini}{Krause
  et~al\mbox{.}}{2014}]%
        {krause2014infuse}
\bibfield{author}{\bibinfo{person}{Josua Krause}, \bibinfo{person}{Adam Perer},
  {and} \bibinfo{person}{Enrico Bertini}.} \bibinfo{year}{2014}\natexlab{}.
\newblock \showarticletitle{INFUSE: interactive feature selection for
  predictive modeling of high dimensional data}.
\newblock \bibinfo{journal}{\emph{IEEE transactions on visualization and
  computer graphics}} \bibinfo{volume}{20}, \bibinfo{number}{12}
  (\bibinfo{year}{2014}), \bibinfo{pages}{1614--1623}.
\newblock


\bibitem[\protect\citeauthoryear{Krishnamurthy, Li, Raghavan, Reiss,
  Vaithyanathan, and Zhu}{Krishnamurthy et~al\mbox{.}}{2009}]%
        {krishnamurthy2009systemt}
\bibfield{author}{\bibinfo{person}{Rajasekar Krishnamurthy},
  \bibinfo{person}{Yunyao Li}, \bibinfo{person}{Sriram Raghavan},
  \bibinfo{person}{Frederick Reiss}, \bibinfo{person}{Shivakumar
  Vaithyanathan}, {and} \bibinfo{person}{Huaiyu Zhu}.}
  \bibinfo{year}{2009}\natexlab{}.
\newblock \showarticletitle{SystemT: a system for declarative information
  extraction}.
\newblock \bibinfo{journal}{\emph{ACM SIGMOD Record}} \bibinfo{volume}{37},
  \bibinfo{number}{4} (\bibinfo{year}{2009}), \bibinfo{pages}{7--13}.
\newblock


\bibitem[\protect\citeauthoryear{Kulesza, Burnett, Wong, and Stumpf}{Kulesza
  et~al\mbox{.}}{2015}]%
        {kulesza2015principles}
\bibfield{author}{\bibinfo{person}{Todd Kulesza}, \bibinfo{person}{Margaret
  Burnett}, \bibinfo{person}{Weng-Keen Wong}, {and} \bibinfo{person}{Simone
  Stumpf}.} \bibinfo{year}{2015}\natexlab{}.
\newblock \showarticletitle{Principles of explanatory debugging to personalize
  interactive machine learning}. In \bibinfo{booktitle}{\emph{Proceedings of
  the 20th international conference on intelligent user interfaces}}.
  \bibinfo{pages}{126--137}.
\newblock


\bibitem[\protect\citeauthoryear{Kulesza, Charles, Caruana, Amershi, and
  Fisher}{Kulesza et~al\mbox{.}}{2019}]%
        {kulesza2019structured}
\bibfield{author}{\bibinfo{person}{Todd Kulesza}, \bibinfo{person}{Denis
  Charles}, \bibinfo{person}{Rich Caruana}, \bibinfo{person}{Saleema~Amin
  Amershi}, {and} \bibinfo{person}{Danyel~Aharon Fisher}.}
  \bibinfo{year}{2019}\natexlab{}.
\newblock \bibinfo{title}{Structured labeling to facilitate concept evolution
  in machine learning}.
\newblock
\newblock
\newblock
\shownote{US Patent 10,318,572.}


\bibitem[\protect\citeauthoryear{Laniado, Eynard, Colombetti,
  et~al\mbox{.}}{Laniado et~al\mbox{.}}{2007}]%
        {laniado2007using}
\bibfield{author}{\bibinfo{person}{David Laniado}, \bibinfo{person}{Davide
  Eynard}, \bibinfo{person}{Marco Colombetti}, {et~al\mbox{.}}}
  \bibinfo{year}{2007}\natexlab{}.
\newblock \showarticletitle{Using WordNet to turn a folksonomy into a hierarchy
  of concepts}. In \bibinfo{booktitle}{\emph{Semantic Web Application and
  Perspectives-Fourth Italian Semantic Web Workshop}}.
  \bibinfo{pages}{192--201}.
\newblock


\bibitem[\protect\citeauthoryear{Manyika, Chui, Miremadi,
  et~al\mbox{.}}{Manyika et~al\mbox{.}}{2017}]%
        {manyika2017future}
\bibfield{author}{\bibinfo{person}{James Manyika}, \bibinfo{person}{Michael
  Chui}, \bibinfo{person}{Mehdi Miremadi}, {et~al\mbox{.}}}
  \bibinfo{year}{2017}\natexlab{}.
\newblock \showarticletitle{A future that works: AI, automation, employment,
  and productivity}.
\newblock \bibinfo{journal}{\emph{McKinsey Global Institute Research, Tech.
  Rep}}  \bibinfo{volume}{60} (\bibinfo{year}{2017}).
\newblock


\bibitem[\protect\citeauthoryear{Mao, Wang, Muller, Varshney, Baldini, Dugan,
  and Mojsilovi{\'c}}{Mao et~al\mbox{.}}{2019}]%
        {mao2019data}
\bibfield{author}{\bibinfo{person}{Yaoli Mao}, \bibinfo{person}{Dakuo Wang},
  \bibinfo{person}{Michael Muller}, \bibinfo{person}{Kush~R Varshney},
  \bibinfo{person}{Ioana Baldini}, \bibinfo{person}{Casey Dugan}, {and}
  \bibinfo{person}{Aleksandra Mojsilovi{\'c}}.}
  \bibinfo{year}{2019}\natexlab{}.
\newblock \showarticletitle{How Data Scientists Work Together With Domain
  Experts in Scientific Collaborations: To Find The Right Answer Or To Ask The
  Right Question?}
\newblock \bibinfo{journal}{\emph{Proceedings of the ACM on Human-Computer
  Interaction}} \bibinfo{volume}{3}, \bibinfo{number}{GROUP}
  (\bibinfo{year}{2019}), \bibinfo{pages}{1--23}.
\newblock


\bibitem[\protect\citeauthoryear{Mitchell, Wu, Zaldivar, Barnes, Vasserman,
  Hutchinson, Spitzer, Raji, and Gebru}{Mitchell et~al\mbox{.}}{2019}]%
        {mitchell2019model}
\bibfield{author}{\bibinfo{person}{Margaret Mitchell}, \bibinfo{person}{Simone
  Wu}, \bibinfo{person}{Andrew Zaldivar}, \bibinfo{person}{Parker Barnes},
  \bibinfo{person}{Lucy Vasserman}, \bibinfo{person}{Ben Hutchinson},
  \bibinfo{person}{Elena Spitzer}, \bibinfo{person}{Inioluwa~Deborah Raji},
  {and} \bibinfo{person}{Timnit Gebru}.} \bibinfo{year}{2019}\natexlab{}.
\newblock \showarticletitle{Model cards for model reporting}. In
  \bibinfo{booktitle}{\emph{Proceedings of the conference on fairness,
  accountability, and transparency}}. \bibinfo{pages}{220--229}.
\newblock


\bibitem[\protect\citeauthoryear{M{\"u}hlbacher, Linhardt, M{\"o}ller, and
  Piringer}{M{\"u}hlbacher et~al\mbox{.}}{2017}]%
        {muhlbacher2017treepod}
\bibfield{author}{\bibinfo{person}{Thomas M{\"u}hlbacher},
  \bibinfo{person}{Lorenz Linhardt}, \bibinfo{person}{Torsten M{\"o}ller},
  {and} \bibinfo{person}{Harald Piringer}.} \bibinfo{year}{2017}\natexlab{}.
\newblock \showarticletitle{Treepod: Sensitivity-aware selection of
  pareto-optimal decision trees}.
\newblock \bibinfo{journal}{\emph{IEEE transactions on visualization and
  computer graphics}} \bibinfo{volume}{24}, \bibinfo{number}{1}
  (\bibinfo{year}{2017}), \bibinfo{pages}{174--183}.
\newblock


\bibitem[\protect\citeauthoryear{Muller, Lange, Wang, Piorkowski, Tsay, Liao,
  Dugan, and Erickson}{Muller et~al\mbox{.}}{2019}]%
        {muller2019datascience}
\bibfield{author}{\bibinfo{person}{Michael Muller}, \bibinfo{person}{Ingrid
  Lange}, \bibinfo{person}{Dakuo Wang}, \bibinfo{person}{David Piorkowski},
  \bibinfo{person}{Jason Tsay}, \bibinfo{person}{Q.~Vera Liao},
  \bibinfo{person}{Casey Dugan}, {and} \bibinfo{person}{Thomas Erickson}.}
  \bibinfo{year}{2019}\natexlab{}.
\newblock \showarticletitle{How Data Science Workers Work with Data: Discovery,
  Capture, Curation, Design, Creation}. In
  \bibinfo{booktitle}{\emph{Proceedings of the 2019 CHI Conference on Human
  Factors in Computing Systems}} (Glasgow, UK) \emph{(\bibinfo{series}{CHI
  '19})}. \bibinfo{publisher}{ACM}, \bibinfo{address}{New York, NY, USA},
  \bibinfo{pages}{Forthcoming}.
\newblock


\bibitem[\protect\citeauthoryear{Nakayama, Kubo, Kamura, Taniguchi, and
  Liang}{Nakayama et~al\mbox{.}}{2018}]%
        {doccano}
\bibfield{author}{\bibinfo{person}{Hiroki Nakayama}, \bibinfo{person}{Takahiro
  Kubo}, \bibinfo{person}{Junya Kamura}, \bibinfo{person}{Yasufumi Taniguchi},
  {and} \bibinfo{person}{Xu Liang}.} \bibinfo{year}{2018}\natexlab{}.
\newblock \bibinfo{title}{{doccano}: Text Annotation Tool for Human}.
\newblock
\newblock
\urldef\tempurl%
\url{https://github.com/doccano/doccano}
\showURL{%
\tempurl}
\newblock
\shownote{Software available from https://github.com/doccano/doccano.}


\bibitem[\protect\citeauthoryear{Nam and Ackerman}{Nam and Ackerman}{2007}]%
        {nam2007arkose}
\bibfield{author}{\bibinfo{person}{Kevin~K Nam} {and} \bibinfo{person}{Mark~S
  Ackerman}.} \bibinfo{year}{2007}\natexlab{}.
\newblock \showarticletitle{Arkose: reusing informal information from online
  discussions}. In \bibinfo{booktitle}{\emph{Proceedings of the 2007
  international ACM conference on Supporting group work}}.
  \bibinfo{pages}{137--146}.
\newblock


\bibitem[\protect\citeauthoryear{Niculescu, Mitchell, and Rao}{Niculescu
  et~al\mbox{.}}{2006}]%
        {niculescu2006bayesian}
\bibfield{author}{\bibinfo{person}{Radu~Stefan Niculescu},
  \bibinfo{person}{Tom~M Mitchell}, {and} \bibinfo{person}{R~Bharat Rao}.}
  \bibinfo{year}{2006}\natexlab{}.
\newblock \showarticletitle{Bayesian network learning with parameter
  constraints}.
\newblock \bibinfo{journal}{\emph{Journal of machine learning research}}
  \bibinfo{volume}{7}, \bibinfo{number}{Jul} (\bibinfo{year}{2006}),
  \bibinfo{pages}{1357--1383}.
\newblock


\bibitem[\protect\citeauthoryear{Passi and Jackson}{Passi and Jackson}{2018}]%
        {passi2018trust}
\bibfield{author}{\bibinfo{person}{Samir Passi} {and} \bibinfo{person}{Steven~J
  Jackson}.} \bibinfo{year}{2018}\natexlab{}.
\newblock \showarticletitle{Trust in data science: collaboration, translation,
  and accountability in corporate data science projects}.
\newblock \bibinfo{journal}{\emph{Proceedings of the ACM on Human-Computer
  Interaction}} \bibinfo{volume}{2}, \bibinfo{number}{CSCW}
  (\bibinfo{year}{2018}), \bibinfo{pages}{1--28}.
\newblock


\bibitem[\protect\citeauthoryear{Pinhanez}{Pinhanez}{2019}]%
        {pinhanez2019machine}
\bibfield{author}{\bibinfo{person}{Claudio Pinhanez}.}
  \bibinfo{year}{2019}\natexlab{}.
\newblock \showarticletitle{Machine Teaching by Domain Experts: Towards More
  Humane, Inclusive, and Intelligent Machine Learning Systems}.
\newblock \bibinfo{journal}{\emph{arXiv preprint arXiv:1908.08931}}
  (\bibinfo{year}{2019}).
\newblock


\bibitem[\protect\citeauthoryear{Piorkowski, Park, Wang, Wang, Muller, and
  Portnoy}{Piorkowski et~al\mbox{.}}{2021}]%
        {piorkowski2021ai}
\bibfield{author}{\bibinfo{person}{David Piorkowski}, \bibinfo{person}{Soya
  Park}, \bibinfo{person}{April~Yi Wang}, \bibinfo{person}{Dakuo Wang},
  \bibinfo{person}{Michael Muller}, {and} \bibinfo{person}{Felix Portnoy}.}
  \bibinfo{year}{2021}\natexlab{}.
\newblock \bibinfo{title}{How AI Developers Overcome Communication Challenges
  in a Multidisciplinary Team: A Case Study}.
\newblock
\newblock
\showeprint[arxiv]{2101.06098}~[cs.CY]


\bibitem[\protect\citeauthoryear{prodigy}{prodigy}{2018}]%
        {prodigy}
\bibfield{author}{\bibinfo{person}{prodigy}.} \bibinfo{year}{2018}\natexlab{}.
\newblock \bibinfo{title}{prodigy}.
\newblock
\newblock
\newblock
\shownote{\url{https://prodi.gy}.}


\bibitem[\protect\citeauthoryear{Raghavan, Madani, and Jones}{Raghavan
  et~al\mbox{.}}{2006}]%
        {raghavan2006active}
\bibfield{author}{\bibinfo{person}{Hema Raghavan}, \bibinfo{person}{Omid
  Madani}, {and} \bibinfo{person}{Rosie Jones}.}
  \bibinfo{year}{2006}\natexlab{}.
\newblock \showarticletitle{Active learning with feedback on features and
  instances}.
\newblock \bibinfo{journal}{\emph{Journal of Machine Learning Research}}
  \bibinfo{volume}{7}, \bibinfo{number}{Aug} (\bibinfo{year}{2006}),
  \bibinfo{pages}{1655--1686}.
\newblock


\bibitem[\protect\citeauthoryear{Ratner, Bach, Varma, and R{\'e}}{Ratner
  et~al\mbox{.}}{2019}]%
        {ratner2019weak}
\bibfield{author}{\bibinfo{person}{Alex Ratner}, \bibinfo{person}{Stephen
  Bach}, \bibinfo{person}{Paroma Varma}, {and} \bibinfo{person}{Chris R{\'e}}.}
  \bibinfo{year}{2019}\natexlab{}.
\newblock \showarticletitle{Weak supervision: the new programming paradigm for
  machine learning}.
\newblock \bibinfo{journal}{\emph{Hazy Research. Available via https://dawn.
  cs. stanford. edu//2017/07/16/weak-supervision/. Accessed}}
  (\bibinfo{year}{2019}), \bibinfo{pages}{05--09}.
\newblock


\bibitem[\protect\citeauthoryear{Ratner, Bach, Ehrenberg, Fries, Wu, and
  R{\'e}}{Ratner et~al\mbox{.}}{2017}]%
        {ratner2017snorkel}
\bibfield{author}{\bibinfo{person}{Alexander Ratner},
  \bibinfo{person}{Stephen~H Bach}, \bibinfo{person}{Henry Ehrenberg},
  \bibinfo{person}{Jason Fries}, \bibinfo{person}{Sen Wu}, {and}
  \bibinfo{person}{Christopher R{\'e}}.} \bibinfo{year}{2017}\natexlab{}.
\newblock \showarticletitle{Snorkel: Rapid training data creation with weak
  supervision}. In \bibinfo{booktitle}{\emph{Proceedings of the VLDB Endowment.
  International Conference on Very Large Data Bases}},
  Vol.~\bibinfo{volume}{11}. NIH Public Access, \bibinfo{pages}{269}.
\newblock


\bibitem[\protect\citeauthoryear{Ratner, De~Sa, Wu, Selsam, and R{\'e}}{Ratner
  et~al\mbox{.}}{2016}]%
        {ratner2016data}
\bibfield{author}{\bibinfo{person}{Alexander~J Ratner},
  \bibinfo{person}{Christopher~M De~Sa}, \bibinfo{person}{Sen Wu},
  \bibinfo{person}{Daniel Selsam}, {and} \bibinfo{person}{Christopher R{\'e}}.}
  \bibinfo{year}{2016}\natexlab{}.
\newblock \showarticletitle{Data programming: Creating large training sets,
  quickly}. In \bibinfo{booktitle}{\emph{Advances in neural information
  processing systems}}. \bibinfo{pages}{3567--3575}.
\newblock


\bibitem[\protect\citeauthoryear{Rule, Drosos, Tabard, and Hollan}{Rule
  et~al\mbox{.}}{2018}]%
        {rule2018aiding}
\bibfield{author}{\bibinfo{person}{Adam Rule}, \bibinfo{person}{Ian Drosos},
  \bibinfo{person}{Aur{\'e}lien Tabard}, {and} \bibinfo{person}{James~D
  Hollan}.} \bibinfo{year}{2018}\natexlab{}.
\newblock \showarticletitle{Aiding collaborative reuse of computational
  notebooks with annotated cell folding}.
\newblock \bibinfo{journal}{\emph{Proceedings of the ACM on Human-Computer
  Interaction}} \bibinfo{volume}{2}, \bibinfo{number}{CSCW}
  (\bibinfo{year}{2018}), \bibinfo{pages}{1--12}.
\newblock


\bibitem[\protect\citeauthoryear{Saleh, Boag, Erdman, and Naumann}{Saleh
  et~al\mbox{.}}{2020}]%
        {saleh2020clinical}
\bibfield{author}{\bibinfo{person}{Shems Saleh}, \bibinfo{person}{William
  Boag}, \bibinfo{person}{Lauren Erdman}, {and} \bibinfo{person}{Tristan
  Naumann}.} \bibinfo{year}{2020}\natexlab{}.
\newblock \showarticletitle{Clinical Collabsheets: 53 Questions to Guide a
  Clinical Collaboration}. In \bibinfo{booktitle}{\emph{Machine Learning for
  Healthcare Conference}}. PMLR, \bibinfo{pages}{783--812}.
\newblock


\bibitem[\protect\citeauthoryear{Schreiber, Schreiber, Akkermans, Anjewierden,
  Shadbolt, de~Hoog, Van~de Velde, Wielinga, Nigel, et~al\mbox{.}}{Schreiber
  et~al\mbox{.}}{2000}]%
        {schreiber2000knowledge}
\bibfield{author}{\bibinfo{person}{A~Th Schreiber}, \bibinfo{person}{Guus
  Schreiber}, \bibinfo{person}{Hans Akkermans}, \bibinfo{person}{Anjo
  Anjewierden}, \bibinfo{person}{Nigel Shadbolt}, \bibinfo{person}{Robert de
  Hoog}, \bibinfo{person}{Walter Van~de Velde}, \bibinfo{person}{Bob Wielinga},
  \bibinfo{person}{R Nigel}, {et~al\mbox{.}}} \bibinfo{year}{2000}\natexlab{}.
\newblock \bibinfo{booktitle}{\emph{Knowledge engineering and management: the
  CommonKADS methodology}}.
\newblock \bibinfo{publisher}{MIT press}.
\newblock


\bibitem[\protect\citeauthoryear{Settles}{Settles}{2009}]%
        {settles2009active}
\bibfield{author}{\bibinfo{person}{Burr Settles}.}
  \bibinfo{year}{2009}\natexlab{}.
\newblock \bibinfo{booktitle}{\emph{Active learning literature survey}}.
\newblock \bibinfo{type}{{T}echnical {R}eport}.
  \bibinfo{institution}{University of Wisconsin-Madison Department of Computer
  Sciences}.
\newblock


\bibitem[\protect\citeauthoryear{Settles}{Settles}{2011}]%
        {settles2011closing}
\bibfield{author}{\bibinfo{person}{Burr Settles}.}
  \bibinfo{year}{2011}\natexlab{}.
\newblock \showarticletitle{Closing the loop: Fast, interactive semi-supervised
  annotation with queries on features and instances}. In
  \bibinfo{booktitle}{\emph{Proceedings of the 2011 Conference on Empirical
  Methods in Natural Language Processing}}. \bibinfo{pages}{1467--1478}.
\newblock


\bibitem[\protect\citeauthoryear{Simard, Amershi, Chickering, Pelton, Ghorashi,
  Meek, Ramos, Suh, Verwey, Wang, et~al\mbox{.}}{Simard et~al\mbox{.}}{2017}]%
        {simard2017machine}
\bibfield{author}{\bibinfo{person}{Patrice~Y Simard}, \bibinfo{person}{Saleema
  Amershi}, \bibinfo{person}{David~M Chickering},
  \bibinfo{person}{Alicia~Edelman Pelton}, \bibinfo{person}{Soroush Ghorashi},
  \bibinfo{person}{Christopher Meek}, \bibinfo{person}{Gonzalo Ramos},
  \bibinfo{person}{Jina Suh}, \bibinfo{person}{Johan Verwey},
  \bibinfo{person}{Mo Wang}, {et~al\mbox{.}}} \bibinfo{year}{2017}\natexlab{}.
\newblock \showarticletitle{Machine teaching: A new paradigm for building
  machine learning systems}.
\newblock \bibinfo{journal}{\emph{arXiv preprint arXiv:1707.06742}}
  (\bibinfo{year}{2017}).
\newblock


\bibitem[\protect\citeauthoryear{Smith, Kumar, Boyd-Graber, Seppi, and
  Findlater}{Smith et~al\mbox{.}}{2018}]%
        {smith2018closing}
\bibfield{author}{\bibinfo{person}{Alison Smith}, \bibinfo{person}{Varun
  Kumar}, \bibinfo{person}{Jordan Boyd-Graber}, \bibinfo{person}{Kevin Seppi},
  {and} \bibinfo{person}{Leah Findlater}.} \bibinfo{year}{2018}\natexlab{}.
\newblock \showarticletitle{Closing the loop: User-centered design and
  evaluation of a human-in-the-loop topic modeling system}. In
  \bibinfo{booktitle}{\emph{23rd International Conference on Intelligent User
  Interfaces}}. \bibinfo{pages}{293--304}.
\newblock


\bibitem[\protect\citeauthoryear{Smith-Renner, Kumar, Boyd-Graber, Seppi, and
  Findlater}{Smith-Renner et~al\mbox{.}}{2020}]%
        {smith2020digging}
\bibfield{author}{\bibinfo{person}{Alison Smith-Renner}, \bibinfo{person}{Varun
  Kumar}, \bibinfo{person}{Jordan Boyd-Graber}, \bibinfo{person}{Kevin Seppi},
  {and} \bibinfo{person}{Leah Findlater}.} \bibinfo{year}{2020}\natexlab{}.
\newblock \showarticletitle{Digging into user control: perceptions of adherence
  and instability in transparent models}. In
  \bibinfo{booktitle}{\emph{Proceedings of the 25th International Conference on
  Intelligent User Interfaces}}. \bibinfo{pages}{519--530}.
\newblock


\bibitem[\protect\citeauthoryear{Snell, Swersky, and Zemel}{Snell
  et~al\mbox{.}}{2017}]%
        {snell2017prototypical}
\bibfield{author}{\bibinfo{person}{Jake Snell}, \bibinfo{person}{Kevin
  Swersky}, {and} \bibinfo{person}{Richard Zemel}.}
  \bibinfo{year}{2017}\natexlab{}.
\newblock \showarticletitle{Prototypical networks for few-shot learning}. In
  \bibinfo{booktitle}{\emph{Advances in neural information processing
  systems}}. \bibinfo{pages}{4077--4087}.
\newblock


\bibitem[\protect\citeauthoryear{Stumpf, Rajaram, Li, Burnett, Dietterich,
  Sullivan, Drummond, and Herlocker}{Stumpf et~al\mbox{.}}{2007}]%
        {stumpf2007toward}
\bibfield{author}{\bibinfo{person}{Simone Stumpf}, \bibinfo{person}{Vidya
  Rajaram}, \bibinfo{person}{Lida Li}, \bibinfo{person}{Margaret Burnett},
  \bibinfo{person}{Thomas Dietterich}, \bibinfo{person}{Erin Sullivan},
  \bibinfo{person}{Russell Drummond}, {and} \bibinfo{person}{Jonathan
  Herlocker}.} \bibinfo{year}{2007}\natexlab{}.
\newblock \showarticletitle{Toward harnessing user feedback for machine
  learning}. In \bibinfo{booktitle}{\emph{Proceedings of the 12th international
  conference on Intelligent user interfaces}}. \bibinfo{pages}{82--91}.
\newblock


\bibitem[\protect\citeauthoryear{Talbot, Lee, Kapoor, and Tan}{Talbot
  et~al\mbox{.}}{2009}]%
        {talbot2009ensemblematrix}
\bibfield{author}{\bibinfo{person}{Justin Talbot}, \bibinfo{person}{Bongshin
  Lee}, \bibinfo{person}{Ashish Kapoor}, {and} \bibinfo{person}{Desney~S Tan}.}
  \bibinfo{year}{2009}\natexlab{}.
\newblock \showarticletitle{EnsembleMatrix: interactive visualization to
  support machine learning with multiple classifiers}. In
  \bibinfo{booktitle}{\emph{Proceedings of the SIGCHI Conference on Human
  Factors in Computing Systems}}. \bibinfo{pages}{1283--1292}.
\newblock


\bibitem[\protect\citeauthoryear{Terveen, Selfridge, and Long}{Terveen
  et~al\mbox{.}}{1995}]%
        {terveen1995living}
\bibfield{author}{\bibinfo{person}{Loren~G Terveen}, \bibinfo{person}{Peter~G
  Selfridge}, {and} \bibinfo{person}{M~David Long}.}
  \bibinfo{year}{1995}\natexlab{}.
\newblock \showarticletitle{Living design memory: framework, implementation,
  lessons learned}.
\newblock \bibinfo{journal}{\emph{Human-Computer Interaction}}
  \bibinfo{volume}{10}, \bibinfo{number}{1} (\bibinfo{year}{1995}),
  \bibinfo{pages}{1--37}.
\newblock


\bibitem[\protect\citeauthoryear{Vale, Lins, and Ferreira}{Vale
  et~al\mbox{.}}{2020}]%
        {Vale20}
\bibfield{author}{\bibinfo{person}{Rafaella Vale},
  \bibinfo{person}{Rafael~Dueire Lins}, {and} \bibinfo{person}{Rafael
  Ferreira}.} \bibinfo{year}{2020}\natexlab{}.
\newblock \showarticletitle{An Assessment of Sentence Simplification Methods in
  Extractive Text Summarization}. In \bibinfo{booktitle}{\emph{Proceedings of
  the ACM Symposium on Document Engineering 2020}} (Virtual Event, CA, USA)
  \emph{(\bibinfo{series}{DocEng '20})}. \bibinfo{publisher}{Association for
  Computing Machinery}, \bibinfo{address}{New York, NY, USA}, Article
  \bibinfo{articleno}{9}, \bibinfo{numpages}{9}~pages.
\newblock
\showISBNx{9781450380003}
\urldef\tempurl%
\url{https://doi.org/10.1145/3395027.3419588}
\showDOI{\tempurl}


\bibitem[\protect\citeauthoryear{Wang, Mittal, Brooks, and Oney}{Wang
  et~al\mbox{.}}{2019}]%
        {wang2019data}
\bibfield{author}{\bibinfo{person}{April~Yi Wang}, \bibinfo{person}{Anant
  Mittal}, \bibinfo{person}{Christopher Brooks}, {and} \bibinfo{person}{Steve
  Oney}.} \bibinfo{year}{2019}\natexlab{}.
\newblock \showarticletitle{How Data Scientists Use Computational Notebooks for
  Real-Time Collaboration}.
\newblock \bibinfo{journal}{\emph{Proceedings of the ACM on Human-Computer
  Interaction}} \bibinfo{volume}{3}, \bibinfo{number}{CSCW}
  (\bibinfo{year}{2019}), \bibinfo{pages}{1--30}.
\newblock


\bibitem[\protect\citeauthoryear{Wang, Wu, Brooks, and Oney}{Wang
  et~al\mbox{.}}{2020}]%
        {wang2020callisto}
\bibfield{author}{\bibinfo{person}{April~Yi Wang}, \bibinfo{person}{Zihan Wu},
  \bibinfo{person}{Christopher Brooks}, {and} \bibinfo{person}{Steve Oney}.}
  \bibinfo{year}{2020}\natexlab{}.
\newblock \showarticletitle{Callisto: Capturing the “Why” by Connecting
  Conversations with Computational Narratives}. In
  \bibinfo{booktitle}{\emph{Proceedings of the 2020 CHI Conference on Human
  Factors in Computing Systems}} \emph{(\bibinfo{series}{CHI '20})}.
  \bibinfo{publisher}{ACM}.
\newblock


\bibitem[\protect\citeauthoryear{Wang, He, and Zhou}{Wang
  et~al\mbox{.}}{2017}]%
        {wang-etal-2017-short}
\bibfield{author}{\bibinfo{person}{Chengyu Wang}, \bibinfo{person}{Xiaofeng
  He}, {and} \bibinfo{person}{Aoying Zhou}.} \bibinfo{year}{2017}\natexlab{}.
\newblock \showarticletitle{A Short Survey on Taxonomy Learning from Text
  Corpora: Issues, Resources and Recent Advances}. In
  \bibinfo{booktitle}{\emph{Proceedings of the 2017 Conference on Empirical
  Methods in Natural Language Processing}}. \bibinfo{publisher}{Association for
  Computational Linguistics}, \bibinfo{address}{Copenhagen, Denmark},
  \bibinfo{pages}{1190--1203}.
\newblock
\urldef\tempurl%
\url{https://doi.org/10.18653/v1/D17-1123}
\showDOI{\tempurl}


\bibitem[\protect\citeauthoryear{Ware, Frank, Holmes, Hall, and Witten}{Ware
  et~al\mbox{.}}{2001}]%
        {ware2001interactive}
\bibfield{author}{\bibinfo{person}{Malcolm Ware}, \bibinfo{person}{Eibe Frank},
  \bibinfo{person}{Geoffrey Holmes}, \bibinfo{person}{Mark Hall}, {and}
  \bibinfo{person}{Ian~H Witten}.} \bibinfo{year}{2001}\natexlab{}.
\newblock \showarticletitle{Interactive machine learning: letting users build
  classifiers}.
\newblock \bibinfo{journal}{\emph{International Journal of Human-Computer
  Studies}} \bibinfo{volume}{55}, \bibinfo{number}{3} (\bibinfo{year}{2001}),
  \bibinfo{pages}{281--292}.
\newblock


\bibitem[\protect\citeauthoryear{Wu, Weld, and Heer}{Wu et~al\mbox{.}}{2019}]%
        {wu2019local}
\bibfield{author}{\bibinfo{person}{Tongshuang Wu}, \bibinfo{person}{Daniel~S
  Weld}, {and} \bibinfo{person}{Jeffrey Heer}.}
  \bibinfo{year}{2019}\natexlab{}.
\newblock \showarticletitle{Local Decision Pitfalls in Interactive Machine
  Learning: An Investigation into Feature Selection in Sentiment Analysis}.
\newblock \bibinfo{journal}{\emph{ACM Transactions on Computer-Human
  Interaction (TOCHI)}} \bibinfo{volume}{26}, \bibinfo{number}{4}
  (\bibinfo{year}{2019}), \bibinfo{pages}{1--27}.
\newblock


\bibitem[\protect\citeauthoryear{Yang, Yuan, and Wang}{Yang
  et~al\mbox{.}}{2019b}]%
        {yang2019knowledge}
\bibfield{author}{\bibinfo{person}{Chi-Lan Yang}, \bibinfo{person}{Chien~Wen
  Yuan}, {and} \bibinfo{person}{Hao-Chuan Wang}.}
  \bibinfo{year}{2019}\natexlab{b}.
\newblock \showarticletitle{When Knowledge Network is Social Network:
  Understanding Collaborative Knowledge Transfer in Workplace}.
\newblock \bibinfo{journal}{\emph{Proceedings of the ACM on Human-Computer
  Interaction}} \bibinfo{volume}{3}, \bibinfo{number}{CSCW}
  (\bibinfo{year}{2019}), \bibinfo{pages}{1--23}.
\newblock


\bibitem[\protect\citeauthoryear{Yang, Steinfeld, and Zimmerman}{Yang
  et~al\mbox{.}}{2019a}]%
        {yang2019unremarkable}
\bibfield{author}{\bibinfo{person}{Qian Yang}, \bibinfo{person}{Aaron
  Steinfeld}, {and} \bibinfo{person}{John Zimmerman}.}
  \bibinfo{year}{2019}\natexlab{a}.
\newblock \showarticletitle{Unremarkable ai: Fitting intelligent decision
  support into critical, clinical decision-making processes}. In
  \bibinfo{booktitle}{\emph{Proceedings of the 2019 CHI Conference on Human
  Factors in Computing Systems}}. \bibinfo{pages}{1--11}.
\newblock


\bibitem[\protect\citeauthoryear{Zhang, Muller, and Wang}{Zhang
  et~al\mbox{.}}{2020}]%
        {zhang2020data}
\bibfield{author}{\bibinfo{person}{Amy~X Zhang}, \bibinfo{person}{Michael
  Muller}, {and} \bibinfo{person}{Dakuo Wang}.}
  \bibinfo{year}{2020}\natexlab{}.
\newblock \showarticletitle{How do Data Science Workers Collaborate? Roles,
  Workflows, and Tools}.
\newblock \bibinfo{journal}{\emph{arXiv preprint arXiv:2001.06684}}
  (\bibinfo{year}{2020}).
\newblock


\bibitem[\protect\citeauthoryear{Zhang, Hu, Deng, Sachan, Yan, and Xing}{Zhang
  et~al\mbox{.}}{2016}]%
        {zhang-etal-2016-learning}
\bibfield{author}{\bibinfo{person}{Hao Zhang}, \bibinfo{person}{Zhiting Hu},
  \bibinfo{person}{Yuntian Deng}, \bibinfo{person}{Mrinmaya Sachan},
  \bibinfo{person}{Zhicheng Yan}, {and} \bibinfo{person}{Eric Xing}.}
  \bibinfo{year}{2016}\natexlab{}.
\newblock \showarticletitle{Learning Concept Taxonomies from Multi-modal Data}.
  In \bibinfo{booktitle}{\emph{Proceedings of the 54th Annual Meeting of the
  Association for Computational Linguistics (Volume 1: Long Papers)}}.
  \bibinfo{publisher}{Association for Computational Linguistics},
  \bibinfo{address}{Berlin, Germany}, \bibinfo{pages}{1791--1801}.
\newblock
\urldef\tempurl%
\url{https://doi.org/10.18653/v1/P16-1169}
\showDOI{\tempurl}


\bibitem[\protect\citeauthoryear{Zhang, Wang, Molino, Li, and Ebert}{Zhang
  et~al\mbox{.}}{2018}]%
        {zhang2018manifold}
\bibfield{author}{\bibinfo{person}{Jiawei Zhang}, \bibinfo{person}{Yang Wang},
  \bibinfo{person}{Piero Molino}, \bibinfo{person}{Lezhi Li}, {and}
  \bibinfo{person}{David~S Ebert}.} \bibinfo{year}{2018}\natexlab{}.
\newblock \showarticletitle{Manifold: A model-agnostic framework for
  interpretation and diagnosis of machine learning models}.
\newblock \bibinfo{journal}{\emph{IEEE transactions on visualization and
  computer graphics}} \bibinfo{volume}{25}, \bibinfo{number}{1}
  (\bibinfo{year}{2018}), \bibinfo{pages}{364--373}.
\newblock


\bibitem[\protect\citeauthoryear{Zhang, Zhao, and LeCun}{Zhang
  et~al\mbox{.}}{2015}]%
        {zhangCharacterlevelConvolutionalNetworks2015}
\bibfield{author}{\bibinfo{person}{Xiang Zhang}, \bibinfo{person}{Junbo Zhao},
  {and} \bibinfo{person}{Yann LeCun}.} \bibinfo{year}{2015}\natexlab{}.
\newblock \showarticletitle{Character-Level { {Convolutional Networks} } for {
  {Text Classification} }}.
\newblock \bibinfo{journal}{\emph{arXiv:1509.01626 [cs]}}
  (\bibinfo{date}{Sept.} \bibinfo{year}{2015}).
\newblock
\showeprint[arxiv]{1509.01626}~[cs]


\bibitem[\protect\citeauthoryear{Zhou, Ackerman, and Zheng}{Zhou
  et~al\mbox{.}}{2011}]%
        {zhou2011cpoe}
\bibfield{author}{\bibinfo{person}{Xiaomu Zhou}, \bibinfo{person}{Mark
  Ackerman}, {and} \bibinfo{person}{Kai Zheng}.}
  \bibinfo{year}{2011}\natexlab{}.
\newblock \showarticletitle{CPOE workarounds, boundary objects, and
  assemblages}. In \bibinfo{booktitle}{\emph{Proceedings of the SIGCHI
  Conference on Human Factors in Computing Systems}}.
  \bibinfo{pages}{3353--3362}.
\newblock


\end{thebibliography}

% \begin{table}[t]
%   \begin{tabular}{p|p|p|p|p|p|p|p}
%   \toprule
%   \textbf{}&\textbf{P1}&\textbf{P2}&\textbf{P3}&\textbf{P4}&\textbf{P5}&\textbf{P6}&\textbf{P7}\\
%         \midrule
%         \textbf{Most favorite}  & Concept Annotation & Perturbation & Concept Annotation%\texttt{data.peek(3)}
%         \\
%         \midrule
%         \textbf{2nd favorite} & Concept Bag of Words & Concept Annotation & Concept Bag of Words%\texttt{data['result'], data['confidence']}
%         \\
%         \midrule
%         \textbf{3rd favorite} & Rule-based (transparency, a few labeled data) & Example-driven conversation (SMEs think aloud labeling data and DS observe it)%\texttt{feature X in data['content']}
%         \\
%         \midrule
%         \textbf{4th favorite} & Random forest (transparency)& Go over analysis together with SMEs (pair authoring~\cite{wang2019data})
%         \\
%         \midrule
%         \textbf{Least favorite} & Watson machine learning library (time) & Example-driven conversation (bug report, new-feature request)  
%         \\
%       \bottomrule
%       \label{tbl:prelim_interviewee}
%          \caption{Data scientist interviewees ranking of different variations}
%     \end{tabular}
   
%     \vspace*{-5mm}
% \end{table}

\end{document}